\documentclass[preprint]{emulateapj}

\newcommand{\beq}{\begin{equation}}
\newcommand{\eeq}{\end{equation}}
\newcommand{\mg}{M_{\rm gas}}
\newcommand{\ms}{M_*}
\newcommand{\msol}{{\rm M_{\odot}}}

\begin{document}

\title{Scaling Relations of Dwarf Galaxies without Supernova-Driven Winds}

\author{Konstantinos Tassis\altaffilmark{1,2},  Andrey V. 
Kravtsov\altaffilmark{1,2,3} \& Nickolay Y. Gnedin\altaffilmark{1,4}}
\altaffiltext{1}{Department of Astronomy and Astrophysics, 
The University of Chicago, Chicago, IL 60637}
\altaffiltext{2}{The Kavli Institute for Cosmological Physics, 
The University of Chicago,
Chicago, IL 60637}
\altaffiltext{3}{Enrico Fermi Institute, The University of Chicago,
Chicago, IL 60637}
\altaffiltext{4}{Particle Astrophysics Group, Fermilab, Batavia, IL 60510}

\begin{abstract}
  Nearby dwarf galaxies exhibit tight correlations between their
  global stellar and dynamical properties, such as circular velocity,
  mass-to-light ratio, stellar mass, surface brightness, and
  metallicity.  Such correlations have often been attributed to gas or
  metal-rich outflows driven by supernova energy feedback to the
  interstellar medium.  We use high-resolution cosmological
  simulations of high-redshift galaxies with and without energy
  feedback, as well as analytic modeling, to
  investigate whether the observed correlations can arise without
  supernova-driven outflows.  We find that the simulated dwarf
  galaxies exhibit correlations similar to those observed as early as
  $z\approx 10$, regardless of whether supernova feedback is included.  We
  also show that the correlations can be well reproduced by our
  analytic model that accounts for realistic gas inflow but
  assumes no outflows, and star formation rate obeying the
  Kennicutt-Schmidt law with a critical density threshold. We argue
  that correlations in simulated galaxies arise due to the
  increasingly inefficient conversion of gas into stars in low-mass
  dwarf galaxies rather than supernova-driven outflows. We also show
  that the decrease of the observed effective yield in low-mass
  objects, often used as an indicator of gas and metal outflows, can be
  reasonably reproduced in our simulations without outflows.  We show
  that this trend can arise if a significant fraction of metals in
  small galaxies is spread to the outer regions of the halo outside
  the stellar extent via mixing. In this case the effective yield can be
  significantly underestimated if only metals within the stellar radius
  are taken into account.  Measurements of gas metallicity in the
  outskirts of gaseous disks of dwarfs would thus provide 
  a key test of such explanation.
\end{abstract}


\maketitle

\section{Introduction}
\label{sec:intro}

Observational studies of galaxies have revealed the existence of tight
correlations between their global stellar and dynamical
properties, such as circular velocity, mass-to-light ratio, stellar
mass, surface brightness, and metallicity. For example, scalings of the
mean metallicity of a galaxy with other global properties such as its
luminosity, stellar mass, or total mass, have been established for
galaxies of a large range of masses and morphologies (e.g.,
\citealp{Leq79, Gar87, Zar94, Garn02, PB02, DW03, Trem04, Pil04, simon_etal06,
lee06}).
Theoretical modeling of these correlations can help us to identify
key physical processes shaping global properties of galaxies.

Supernova energy feedback to the interstellar medium (ISM) with
putative associated gas outflows from galaxies has long been a favored
mechanism to explain these correlations and other properties of
low-mass galaxies \citep[e.g.,][]{Lar74,DS86,arimoto_yoshii87}.  The
strong effect of the SN feedback on properties of low-mass galaxies is
also a standard assumption of the semi-analytic models of galaxy
formation \citep[e.g.,][]{lacey_etal93,KWG93,cole_etal94,somerville_primack99,
benson_etal03,croton_etal06}.
\citet{DW03} argue, for example, that the correlations between
circular velocity, stellar mass, metallicity, and stellar surface
densities they find for the nearby dwarf galaxies can be reproduced
with a simple semi-analytic model incorporating the effect of
supernova feedback on the gas component of dwarf galaxies.

The existence of galactic winds has indeed been observationally
established in star-forming galaxies at low and high redshifts
\citep[e.g.,][see \citeauthor{veilleux_etal05}
\citeyear{veilleux_etal05} for a recent
review]{Hetal00,Pett01,Metal02,strickland_etal04,martin05,ott_etal05}. 
An indirect evidence for metal loss in winds is the high-metallicity 
of the diffuse intergalactic medium in groups and clusters 
\citep[e.g.,][]{renzini_etal93}.
However, the extent to which the outflows affect the global properties of
galaxies and whether the outflow gas escapes the gravitational
potential of its host halo or rains back down to the disk remains
uncertain. \citet{maclow_ferrara99} and \citet{dercole_brighenti99}
used numerical simulations to show that global gas blowout is
inefficient in all but the smallest mass halos \citep[$M\lesssim
10^7{\rm\ M_{\odot}}$; see also][]{ferrara_tolstoy00,marcolini_etal06}, although
metal-enriched SN ejecta may be removed efficiently. It is also
uncertain whether correlations of metallicity with galaxy stellar mass
and mass-to-light ratios can be attributed solely to such winds.

In the context of cosmological simulations, the SN feedback is usually
ineffective, unless \emph{ad hoc} phenomenological recipes enhancing
SN feedback are employed \citep[e.g.,][]{navarro_white93,navarro_steinmetz97,
thacker_couchman01,marri_white03,springel_hernquist03,scannapieco_etal06}.
It is likely that this inefficiency is at least partially due to the
inability of current cosmological simulations to resolve the scales
and processes relevant to stellar feedback.  However, it is also
possible that the actual effects of feedback on the global properties
of galaxies are fairly small in reality.  Cosmological simulations at
this point cannot make \emph{ab initio} predictions about the importance
or inefficiency of stellar feedback.  Observations, on the other hand, although
showing evidence for the presence of large-scale winds in actively starforming
galaxies \citep[e.g., see][for review]{veilleux_etal05},  
are rather uncertain in their estimates of wind mass loss 
to provide reliable direct observationally-motivated 
feedback recipes for implementation in simulations. Nevertheless, we can gauge the
importance of supernova energy injection by comparing galaxies formed in
simulations performed with different assumptions about feedback to
observations.  In this respect, the smallest dwarf galaxies present
the ideal test case because they can be expected to be the most
susceptible to the effects of SN feedback due to their shallow
potential wells \citep{Lar74,DS86}.

The role of feedback in the formation of dwarf galaxies has been
investigated in several recent studies.  \cite{Tetal03} used Eulerian
AMR simulations of galaxy formation with star formation and SN
feedback and found that correlations between mass-to-light ratio and
metallicity and stellar mass and metallicity, similar to those
observed for the nearby dwarfs, are exhibited by dwarf galaxies in
their simulations at $z \gtrsim 3$.  The simulations of
\citet{Tetal03} used a rather extreme amount of energy per supernovae
to maximize the effects of feedback. They did not test, however, whether
feedback or some other mechanism is in fact the dominant factor in
shaping the correlations.  More recently, \cite[][]{Kob06} found a
tight relation between stellar metallicity and stellar mass at all
redshifts in their Smooth Particle Hydrodynamics (SPH) cosmological galaxy
formation simulations, which they attributed to the mass-dependent
galactic winds. The winds in the model of \citet[][]{Kob06} do affect the
gas and metal content of galaxies significantly, with the effect
increasing towards lower mass systems (see their Fig.~16), which plays
an important role in shaping the resulting correlation of metallicity
and stellar mass. 

\citet{derossi_etal06} have recently studied the
origin of luminosity- and stellar mass-metallicity relations in
cosmological simulations. They find that correlations similar to those
observed arise already at high redshifts, although supernova feedback
in their simulations is not efficient. This is consistent with our
results presented below. \cite{simon_etal06} seeked observational correlations
between metallicity, surface brightness and mass-to-light ratio within a
single galaxy (M33) and found that indeed such correlations
exist. They used SPH simulations and 
analytical calculations to show that feedback is not responsible 
for these correlations. They found that the presence of feedback decreases the
scatter in such relations, although this may at least in part be due
to their numerical implementation of feedback.
Most recently, \citet{brooks_etal06} 
presented a study of the origin and evolution of the mass-metallicity
relation in cosmological simulations. They also concluded that
the relation arises not due to the mass loss in winds, but 
due to increasing inefficiency of star formation in smaller
mass galaxies. However, they argue that 
low effective yields of dwarf galaxies can only be
explained by the loss of metals in winds. We argue below that 
there is an alternative explanation for the low effective yields
of dwarfs. 

In this paper we investigate the origin of the observed mass-metallicity 
and other correlations
of dwarf galaxies with the specific goal of testing the role of SN feedback. In
particular, we discuss results of several high-resolution
cosmological simulations of galaxy formation started from the same
initial conditions and run with the same prescriptions for star
formation and metal enrichment of the ISM, but with different models
for gas cooling, UV heating (optically thin approximation vs.
self-consistent radiative transfer of the UV radiation), and supernova
feedback (see \S~\ref{sims} for details). We follow the simulations
until $z\sim 3$.

Although our models simulate galaxies at high redshifts, we study the
evolution of the correlations with redshift from $z=9$ to $z\sim 3$
and show that the trends with metallicity and stellar mass are
approximately preserved as the galaxy population evolves allowing
extrapolation of the results to the present epoch. Therefore, our
results and conclusions give us insight to the processes responsible
for the properties of observed low-redshift dwarfs. There is indeed
observational indication that the stellar mass-metallicity
correlation, for example, is already established by $z\sim 1-2$
\citep{kobulnicky_kewley04,savaglio_etal05,Erb06}, although the metallicities of
galaxies of a given stellar mass appear to be lower by $\approx
0.3$~dex at $z\sim 2$ compared to $z=0$ \citep{savaglio_etal05,Erb06}. In addition,
for some of the Local Group dwarf galaxies that exhibit these
correlations, a significant fraction of their stellar mass was built at
$z\gtrsim 3$ \citep[e.g.,][]{dolphin_etal05}.  The correlations for
these dwarfs should thus be largely set at high redshifts, which we
probe with our simulations.

Our main result is that correlations similar to those observed can be
reproduced in all runs, which suggests that supernova energy feedback
is not required to explain the observed properties of dwarfs.  To
interpret the simulation results we use a simple open-box analytic
model for the evolution of the baryonic component of dwarf galaxies.
We show that a model with gas accretion but no
outflows and the Kennicutt law for star formation with the critical
density threshold for star formation reproduces the results of our
simulations.

This paper is organized as follows. In \S \ref{obs} we summarize the 
observationally established correlations between global 
properties of dwarf galaxies that we wish to interpret. 
Our analytic model and chemical evolution calculations are described in \S 
\ref{ce}. In \S \ref{sims} we describe in
detail our cosmological simulations, and in \S \ref{res} we discuss
how global properties of the simulated galaxies are correlated with
each other, and we compare these correlations to the observed ones,
and to the analytic model of \S \ref{ce}.  A
discussion of observational results on the effective
yield and corresponding constraints on the existence of outflows 
is given in \S \ref{syeff}.  We discuss our findings in \S
\ref{disc}, and we summarize our conclusions in \S \ref{sum}.

\section{Observed Correlations}\label{obs}

Correlations of observables in dwarf galaxies relate 
the stellar mass, $\ms$, with 
the metallicity $Z$, the surface brightness $\mu_*$ and the maximum 
circular velocity 
$V_{\rm m}$ of dwarf galaxies. These can be parametrized as the power
law relations: 
\begin{equation}
Z \propto \ms ^{n_Z}\,;\,\,\, \mu_* \propto \ms ^{n_\mu}\,;\,\,\,
V_{\rm m}\propto \ms^{n_V}\,.
\end{equation}
The existence of these correlations (in several cases in the form of
correlations between absolute magnitudes, rather than stellar masses,
and $Z$, $\mu_*$ and $V_{\rm m}$) has been
established through observations of galaxies with a large range of
masses and in a variety of settings, including nearby dwarfs 
(e.g. \citealp{Skil89, DW03} based on observations taken from 
\citealp{mateo98},\citealp{vdb00} and references therein; 
\citealp{lee06,vanzee_haynes06}), dwarf galaxies 
in the  Sloan Digital Sky Survey
\citep[SDSS;][]{Trem04, Kauf03, Blan03, Bern03}, the Hubble Deep Field
\citep{Driv99}, the Ursa Major cluster \citep{BDJ01} 
and the Virgo cluster \citep{FB94}. However, the exact values of the
slopes of these relations are affected by systematic uncertainties
involved in obtaining stellar masses from the observed quantities
(magnitudes). In general, the slopes depend on the
the mass range in which they are determined. 

For example, in the case of the $Z-\ms$ scaling, \cite{DW03} find that
for the Local Group dwarfs in the mass range of $10^6 \msol \lesssim \ms
\lesssim 10^{10} \msol$, $n_Z \sim 0.4$, while the SDSS data of
\cite{Trem04} (see also \citealp{gallazzi_etal05}) indicate that $n_Z$
increases with decreasing $\ms$: it is quite flat for $\ms \gtrsim 3
\times 10^{10} \msol$, while it appreciably exceeds the \cite{DW03}
value by $M\sim {\rm few}\times 10^8 \msol$. Most recently,
\citet{lee06} find $n_Z\approx 0.29\pm 0.03$ for the dwarf galaxies in their
sample.

In the case of the
$\mu_*-\ms$ scaling, \cite{DW03} find for the Local Group dwarfs 
the slope of $n_\mu\sim 0.55$, although their data points exhibit
a scatter around the best-fit line which is large compared to their
adopted error bars. For the same scaling, \cite{Kauf03} using SDSS data
for the surface density of stellar mass within the half-light radius,
find that for a given value of $\ms$ there is appreciable scatter
(about an order of magnitude) in $\mu_*$, and the {\em median} value
of $\mu_*$ scales roughly as $\ms^{0.63}$ for 
$10^8 \lesssim \ms \lesssim 3 \times 10^{10}
\msol$, and it becomes almost independent of $\ms$ for larger $\ms$. 

In the case of the $\ms-V_{\rm m}$ scaling (the Tully-Fisher relation),
 \cite{BDJ01} using galaxies with $3 \times 10^8 \msol \lesssim \ms
\lesssim 10^{11} \msol$ find $n_V \sim 0.21-0.24$, and a shallower
correlation with {\em baryonic} mass (i.e. stars plus gas), 
$n_V \sim 0.27-0.30$. 
Several recent studies also find that the $V_{\rm m} - \ms$ 
relation for
low-mass galaxies ($V_{\rm m} \lesssim 90$~km/s) is steeper and exhibits
considerably more scatter than the baryonic Tully-Fisher
relation of the same galaxies 
\citep{matthews_etal98,mcgaugh_etal00,BDJ01,
gurovich_etal04,mcgaugh05,geha_etal06}.
These studies find slopes of $n_V\approx 0.1-0.2$ 
for the $\ms-V_{\rm m}$ relation
of the dwarf galaxies \citep{mcgaugh_etal00,gurovich_etal04} 
and $n_V\approx 0.25-0.33$ for the baryonic $M_{\rm baryon}-V_{\rm m}$ 
relation 
\citep[e.g.,][]{mcgaugh_etal00,verheijen01,BDJ01,mcgaugh05,geha_etal06}. 
For the Local Group dwarfs, \cite{DW03}
obtain $n_V \sim 0.37$ for $10^8 \msol
\lesssim \ms \lesssim 10^{10} \msol$ and $n_V \sim 0 $ for $10^6 \msol
\lesssim \ms \lesssim 10^8 \msol$. 

Finally, another tight relation found in Local Group dwarf galaxies is the
correlation between the dynamical mass-to-light ratio and average metallicity 
\citep{PB02}:
\begin{equation}
\log (M_{\rm dyn}/L_V) \propto - {\rm [Fe/H]}\,.
\end{equation}

\section{Analytical Model for the Evolution of the 
Baryonic Component of Dwarf Galaxies} 
\label{ce}

In this section we present a simple analytic model 
for the evolution of the baryonic component 
(gas, stars and metals) of dwarf galaxies at high redshifts, 
which we will make to interpret our simulation results. 
The model is based on the following assumptions:

(1) Galaxies accrete gas at a rate consistent with simulations, but  no gas or metals escape outside the virial radius
of the parent halo.

(2) The gas mass of each galaxy is proportional to its total (dynamical) mass, $\mg \propto M$. At early epochs
$ \ms \ll \mg$, which implies for the baryonic mass $M_{\rm b} = \mg +\ms \approx \mg$.  
The gas fraction
within halos in our simulations is approximately independent of the
total (dynamical) mass for $M_{\rm tot} \gtrsim 10^9 \msol$ (see  \S \ref{res}, Fig. \ref{MgMtot}), which is
consistent with the results of independent simulations of
\citet{Tetal03}. Therefore this assumption is an excellent
approximation for all but the smallest objects in the simulation,
which are excluded in our analysis of other global quantities.
Although at lower redshifts the total baryon fraction in small mass
galaxies $M<10^{10}\ M_{\odot}$ can be suppressed due to heating by
the cosmic UV background \citep[e.g.,][]{Gn00,hoeft_etal06}, our
results would still be applicable to galaxies in which a significant
fraction of stars is formed at high redshifts before such suppression
has occurred in the mass range we consider.  We should stress that
we intend to apply our model only to such stellar populations.

(3) The gas in galaxies forms an exponential disk with radial extent
proportional to the virial radius of the parent halo. The
disk is assumed to be more extended and consequently less compressed
for smaller objects with $T_{\rm vir}<10^4$~K, in which gas cannot
cool efficiently.

(4) Gas is
converted into stars at a rate given by the Kennicutt law of star
formation with a density threshold
\citep[e.g.,][]{martin_kennicutt01}. Gas at densities below the
threshold does not form stars. Note that the Kennicutt law with the
threshold is reproduced by our numerical simulations, but it is {\em
not} assumed {\it a priori}. The numerical simulations implement a
much simpler recipe for star formation (see \S\ref{sims}). 

(5) Instantaneous recycling approximation: i.e., no time delay is assumed between the birth of a generation of stars and
the corresponding metal enrichment of the ISM. Hence, we implicitly
assume that the metal enrichment due to type II supernovae is
dominant. This is appropriate for all but the most massive dwarf
galaxies.

The key ingredient of our model is the relation between the gas mass and
the stellar mass of a galaxy. We will first derive this correlation,
and subsequently use it to derive relations between the stellar mass,
gas mass, stellar surface density, metallicity, and circular velocity
of high-$z$ galaxies.

\subsection{Stellar Mass - Gas Mass Relation} \label{secmstmg}

The relation between stellar mass and total gas mass in dwarf galaxies is
central to understanding the origin of correlations between other
observable quantities\footnote{This relation may not hold
for some galaxies, if, for instance, their gas is lost due to tidal or
ram pressure stripping.}. In this section, we derive such a
relation. In subsequent sections, we will use it to explain the
correlations discussed in \S \ref{obs} and we will directly compare it
against our simulation results.

The rate of change of the gas mass, assuming 
that there are no significant gas outflows due to SN-driven 
winds, is
\beq\label{gas}
\frac{d}{dt}(\mg) = F-a \psi\,,
\eeq
where $F$ is the gas mass accretion rate, $\psi$ is the star formation
rate, and $a$ is the fraction of stellar mass locked in long-lived stars and 
remnants (i.e., not returned to the ISM). The rate of change 
of the stellar mass is
\beq \label{diffstM}
\frac{d\ms}{dt} = a \psi\,.
\eeq 
If the 
dependence of $\psi$ and $F$ on $\mg$, as 
well as any explicit dependence on redshift, are known, 
then equations~(\ref{gas}) and~(\ref{diffstM}) can be 
integrated in time as a system of two ordinary differential 
equations. Different initial conditions for $\mg$ at some
early time (for which we can safely assume that $\ms = 0$) 
will produce different pairs of $(\mg,\ms)$
at any time of interest. In this way, we calculate 
parametrically the prediction of our analytic 
model for the $\ms-\mg$ relation. 

As we show in the Appendix, $F$ is proportional to the gas mass and
$\psi$ depends on the extent of the gaseous disk (which in turn
depends on the dynamical mass of the galaxy and the angular momentum
parameter, $\lambda$) and the star formation density threshold,
$\Sigma_{\rm th}$.  

\begin{figure}
\plotone{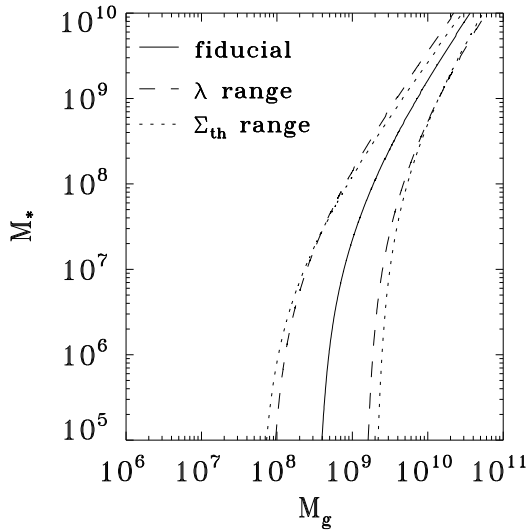}
\caption{\label{range}  Sensitivity of the analytic model to 
the angular momentum parameter $\lambda$ and the surface density threshold
for star formation, $\Sigma_{\rm th}$, at $z=4$. Solid line: fiducial 
model with $\lambda = 0.033$ (its most probable value) and 
$\Sigma_{\rm th} = 5 \, \msol {\, \rm pc^{-2} }$. Dashed lines: 
model results for $\Sigma_{\rm th} = 5 \, \msol {\, \rm pc^{-2} }$
and $\lambda = 0.023$ (left dashed line) or $\lambda = 0.044$ 
(right dashed line). Dotted lines: model results for $\lambda = 0.033$ 
and $\Sigma_{\rm th} = 2 \, \msol {\, \rm pc^{-2} }$ (left dotted line)
or $\Sigma_{\rm th} = 10 \, \msol {\, \rm pc^{-2} }$ (right dotted line).
}
\vspace{0.5cm}
\end{figure}

The free parameters which control the normalization and the 
shape of the $\mg - \ms$ relation 
are $\lambda$ and $\Sigma_{\rm th}$.
The solution of Eqs. (\ref{gas}) and (\ref{diffstM}) for the fiducial values of these parameters is plotted with the solid line in Fig. \ref{range}. At high gas masses, $\ms$ has a power-law dependence on $\mg$, with a slope set by the slope of the star-formation law. At lower $\mg$, this relation exhibits a break, which is a result of the threshold in the star-formation law resulting in very inefficient star formation at the low densities achieved by inefficiently cooling low-mass galaxies.

Remarkably, we find that the best fit to our data at $z=4$ is obtained
 when both free parameters have their fiducial values, $\lambda =
 0.033$ (the most probable value) and $\Sigma_{\rm th} = 5 \, \msol
 {\, \rm pc^{-2} }$ (the ``canonical'' observational
 value). Figure~\ref{range} shows the sensitivity of our analytic
 model to the values of $\lambda$ and $\Sigma_{\rm th}$.  The solid
 line represents the $z=4$ model results for the fiducial parameter
 values (which are also the best-fit parameters). The dashed lines
 show the range of $\lambda$ values that brackets the scatter of our
 simulation data points at $z=4$. The range of $\lambda$ values
 represented here is $0.023 \lesssim \lambda \lesssim 0.044$ (which
 contains $\sim 40 \%$ of the area under the $\lambda-$probability
 distribution, and is centered around the distribution peak). In this
 case, $\Sigma_{\rm th}$ is kept at its fiducial value. The dotted
 lines show the range of $\Sigma_{\rm th}$ values that brackets the
 scatter of our data.  The range of $\Sigma_{\rm th}$ values
 represented here is $2 \, \msol \, {\rm pc ^{-2}} \lesssim
 \Sigma_{\rm th} \lesssim 10 \, \msol \, {\rm pc ^{-2}}$. In this
 case, $\lambda$ is kept at its fiducial value.  The asymptotic slope
 of the curve at higher masses, remains largely independent of
 parameter choice.

\subsection{Mass-to-light ratio vs. metallicity}
\label{mlzmodel}

Under the assumptions of negligible inflow or outflow of metals
and instantaneous recycling, and for a fixed initial mass function, 
the mass of metals of low-$Z$, high-$z$ galaxies 
is proportional to the stellar mass\footnote{For this proportionality
to hold, $Z$ needs to be low enough so that the mass of freshly
produced metals is much larger than the mass of metals locked in long-lived
stars and stellar remnants. This can be shown to be true for $Z\ll
Z_{\odot}$ for the metallicity enrichment implemented in our
simulations, and which always holds in our simulated galaxies.}.
From $M_{\rm Z} = Z\mg \propto \ms$ we can immediately write 
\beq\label{zmtl0}
Z  \propto \frac{\ms}{\mg}\,.
\eeq 
Assumption (2),  $\mg \propto M$, then gives $Z\propto\ms/M$. 
The luminosity of an object, $L$, can be considered proportional
to its stellar mass\footnote{This approximation is appropriate for older
stellar populations. We use it here because we assume that the stellar
populations in our high-$z$ galaxies would actually be observed today,
when they have aged substantially.}. Then, Eq. (\ref{zmtl0})
becomes 
$Z \propto L/M$ or
\beq\label{zmtl}
M/L \propto Z^{-1},
\eeq
in agreement with observations and, as shown in \S \ref{res}, 
 the results of our simulations. 
This is a simple interpretation of the correlation between the
mass-to-light ratio and metallicity found by \cite{PB02}.

\subsection{Additional Correlations}

We can use the $\mg-\ms$ relation we derived in \S \ref{secmstmg}
in combination with Eq. (\ref{zmtl0}) and assumptions $\mg\propto M$ and $L \propto \ms$ to  
derive additional relations between stellar mass and 
stellar surface density, metallicity, and circular velocity
of high-$z$ galaxies. We will overplot these relations with our
simulation results and discuss them in \S \ref{res}.
\section{Numerical Simulations}\label{sims}

\begin{deluxetable*}{lccc}
\tablecaption{Simulations\label{table1}}
\tablewidth{0pt}
\tablehead{
\colhead{Simulation\tablenotemark{a}} & \colhead{SN energy feedback} & \colhead{cooling} & 
\colhead{3-D rad. transfer}
}
\startdata\\
FEC    & yes & equilibrium     & no\\
NFEC & no  & equilibrium     & no \\
FNEC-RT & yes & non-equilibrium & yes\\
F2NEC-RT& yes & non-equilibrium & yes\\
\enddata
\tablenotetext{a}{
Throughout the text
we  denote different runs with abbreviations indicating the
different implementations of key physical processes. In
the abbreviations: F is for feedback, NF is for simulations
with SN feedback explicitly turned off, and F2 is for an alternative
implementation of feedback with delay after the star formation event; EC is for equilibrium cooling
using Cloudy tables, and NEC is for non-equilibrium cooling and
reaction network of ionic species of H, He, and H$_2$; RT indicates runs
with self-consistent radiative transfer.}
\end{deluxetable*}

In this study we use four simulations (Table \ref{table1}) of the
early ($z\gtrsim 3$) stages of evolution for a Lagrangian region of a
MW-sized system of total (dark matter and baryons) virial mass
$\approx 10^{12}h^{-1}\ \rm M_{\odot}$ at $z=0$. The simulations were
performed using the Eulerian, gasdynamics $+N$-body Adaptive
Refinement Tree (ART) code \citep{ketal97,Kth}.  In both the
gasdynamics and gravity calculations, a large dynamic range is
achieved through the use of adaptive mesh refinement (AMR).  At the
analyzed epochs, the galaxy has already built up a significant portion
of its final total mass: $1.3\times 10^{10}h^{-1}\ \rm M_{\odot}$ at
$z=9$ and $2\times 10^{11}h^{-1}\ \rm M_{\odot}$ at $z=4$. The
evolution is started from a random realization of a gaussian density
field at $z=50$ in a periodic box of $6h^{-1}\ \rm Mpc$ with an
appropriate power spectrum and is followed assuming flat $\Lambda$CDM
model: $\Omega_0=1-\Omega_{\Lambda}=0.3$, $\Omega_b=0.043$,
$h=H_0/100=0.7$, $n_s=1$, and $\sigma_8=0.9$.

To achieve the high mass resolution, a low-resolution simulation was
run first and a galactic-mass halo was selected. A Lagrangian region
corresponding to five virial radii of the object at $z=0$,
corresponding to a region of $\sim 3h^{-1}$~ comoving Mpc in diameter,
was then identified at $z=50$ and re-sampled with additional
small-scale waves \citep[][the specific procedure described here is
described in \citeauthor{klypin_etal01}
\citeyear{klypin_etal01}]{navarro_white94}.  The total number of DM
particles in the high-resolution Lagrangian region is $2.64\times
10^6$ and their mass is $m_{\rm DM}=9.18\times 10^5h^{-1}\ \rm
M_{\odot}$. In addition to the main progenitor of the MW-sized system
this Lagrangian region contains several dozens of smaller galaxies
spanning a wide range of masses. We will use these galaxies to analyze
correlations between their properties.

The code used a uniform $64^3$ grid to cover the entire computational
box. The Lagrangian region, however, was always unconditionally
refined to the third refinement level, corresponding to the effective
grid size of $512^3$. As the matter distribution evolves, the code
adaptively and recursively refines the mesh in high-density regions
beyond the third level using two refinement criteria: gas and DM mass
in each cell. A mesh cell was tagged for refinement if its gas {\em
or} DM mass exceeded $1.2\times 10^6h^{-1}\ \rm M_{\odot}$ and
$3.7\times 10^6h^{-1}\ \rm M_{\odot}$, respectively.  The maximum
allowed refinement level was $l_{\rm max}=9$.  The volume of
high-density cold star forming disks forming in DM halos was refined
to $l_{\rm max}=9$. The physical size of mesh cells in the simulations
was $\Delta x_l=26.16\,\,[10/(1+z)] 2^{9-l}\ \rm pc$, where $l$ is the
cell's level of refinement.  Each refinement level was integrated with
its own time step $\Delta t_l=\Delta t_02^{-l}\approx 2\times 10^4\,\,
2^{9-l}\ \rm yrs$. The time steps are set by a global Courant-Friedrich-Levy
 condition.

The analyzed simulations were started from the same initial
conditions, but evolved using different assumptions about cooling and
heating processes accompanying galaxy formation. All four runs
include star formation using the recipe described in \citet{K03}. Namely,
the gas is converted into stars on a characteristic gas consumption
time scale, $\tau_{\ast}$: $\dot{\rho}_{\ast}=\rho_{\rm
  g}/\tau_{\ast}$. We use constant, density-independent $\tau_{\ast}$.
This is different from the commonly used density-dependent efficiency,
but may be more appropriate for star forming regions on $\sim 100$~pc
scales which are resolved in the simulations where constant star
formation efficiency is indicated by observations
\citep[e.g.,][]{young_etal96,wong_blitz02}. As shown by \citet{K03},
such constant efficiency assumption at the scales of molecular clouds
results in the Kennicutt-like star formation law on kiloparsec scales,
although a Kennicutt law {\em is not a priori assumed}.
The star formation was allowed to take place only in the densest
cold regions, $\rho_{\rm g} > \rho_{\rm SF}$ and $T_{\rm g}<T_{\rm SF}$, 
but no other criteria (like
the collapse condition $\nabla\cdot {\bf v} < 0$) were imposed. We
used $\tau_{\ast}=4$~Gyrs, $T_{\rm SF}=10^4$~K, and $\rho_{\rm SF}=1.64{\ \rm
  M_{\odot}pc^{-3}}$ corresponding to atomic hydrogen number density of $n_{\rm
  H, SF}=50{\ \rm cm^{-3}}$.

Our fiducial simulation (FEC) includes metallicity-dependent cooling
and UV heating due to cosmological ionizing background with
equilibrium cooling and heating rates tabulated for the temperature
range $10^2<T<10^9$~K using the {\tt Cloudy} code \citep[ver.
96b4,][]{ferland_etal98}. The cooling and heating rates take into
account Compton heating/cooling of plasma, UV heating, atomic and
molecular cooling. It also includes energy feedback and chemical
enrichment due to supernovae assuming the \citet{MS79}
stellar initial mass function (IMF) and stellar masses in the range
$0.1-100\ \rm M_{\odot}$. All stars with $M_{\ast}>8{\ \rm M_{\odot}}$
deposit $2\times 10^{51}$~ergs of thermal energy and a mass $f_{\rm
  Z}M_{\ast}$ of heavy elements in their parent cell (no delay of
cooling was introduced in these cells).  The metal fraction is $f_{\rm
  Z}= {\rm min}(0.2,0.01M_{\ast}-0.06)$, which crudely approximates
the results of \citet{woosley_weaver95}.

The second simulation, NFEC, is identical to the FEC run in all
respects, except that it did not include the energy injection due to
SNe (chemical enrichment due to SNe is still included).  Note that the
cooling rates in the run with feedback accounted for the local
metallicity of the gas, while in the run with no feedback the
significantly lower zero-metallicity cooling rates were used\footnote{
This choice does not appreciably affect the star-formation properties of 
small galaxies, as shown by Kravtsov (2003), since, at least for the 
densities we are resolving here, the density probability distribution function  
does not change significantly.}.
  
The third simulation, FNEC-RT, includes star formation and SN
enrichment and energy feedback in the same way as the FEC run, but
uses self-consistent 3-D radiative transfer of UV radiation from
individual stellar particles using the OTVET algorithm \citep{ga01,icam06}
and follows non-equilibrium chemical network of hydrogen and helium
species (the details of the specific implementation of the OTVET
algorithm on adaptively refined meshes will be described
elsewhere). The simulation thus includes non-equilibrium cooling and   
heating rates which make use of the local abundance of atomic,
molecular, and ionic species and UV intensity followed
self-consistently during the course of the simulation. The
metallicity-dependence of cooling rates is taken into account using
optically thin equilibrium metal cooling functions from \citet{sd93}
in the high metallicity regime and from \citet{p70} and \citet{dm72} in the
low metallicity regime. The molecular hydrogen formation on dust is
taken into account using \citet{cs04} rates, and gas cooling on
dust is based on \citet{d81}.

A severe limitation of our current treatment of molecular hydrogen
formation is that self-shielding of molecular lines in the
Lyman-Werner band is not taken into account, which results in a
substantial underestimate of the molecular fractions at the high density
regime ($n\ga100{\rm\,cm}^{-3}$). Note, however, that since our star
formation criterion is based on the total rather than molecular gas
density, this does not affect our star formation rates significantly.

Finally, the fourth simulation, F2NEC-RT, is identical to the FNEC-RT
but SN energy is injected with a delay of $10^7$~years after the star
forming event and energy release is spread over $10^7$~years after
that.  This variation is intended to test the sensitivity of our
results to the specific details of the feedback implementation. Energy
injection with delay was argued to enhance the efficiency and impact
of the SN feedback on the ISM \citep{slyz_etal05}, as the stellar
particles have time to leave the densest regions where they form and
release their energy at lower densities where the cooling time is
longer. This change of feedback recipe results in a significant
enhancement of the star formation efficiency. Perhaps surprisingly, the
F2NEC-RT simulation forms nearly twice as much stellar mass as the
simulation FNEC-RT. The reason is simple. With a delayed energy 
release the star forming regions are generally left intact and 
continuing forming stars in the simulation F2NEC-RT, while in the
FNEC-RT the temperature of gas in star forming regions is increased
due to SN feedback and star formation ceases for some period of time. 

We show below that the details of the specific implementation of 
feedback in our simulations do not affect the correlations
of galaxy properties and our conclusions.

\section{Results}\label{res}
\begin{figure*}[h]
\plotone{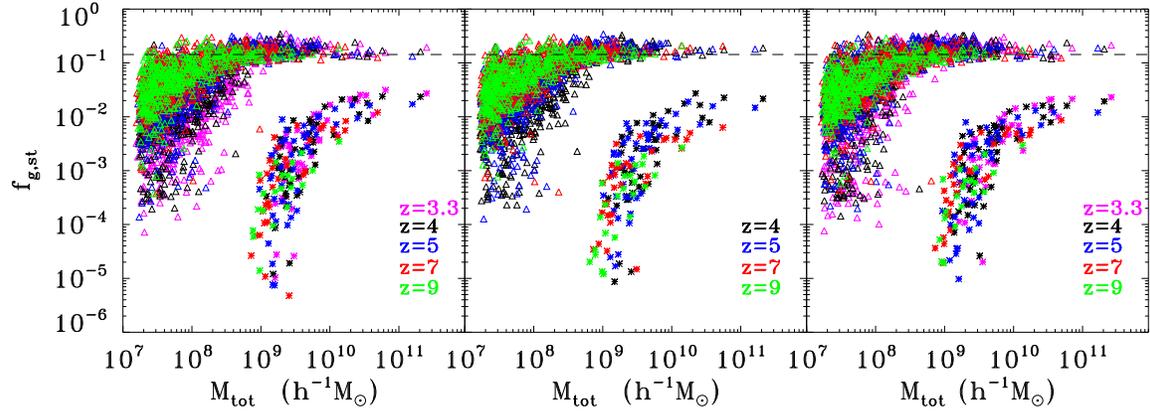}
\caption{\label{MgMtot}  Gas and stellar mass fractions of objects in our
simulations plotted against the total (dark matter + total baryonic) 
mass of each object. 
Left panel: simulation FEC; middle panel: simulation NFEC; right panel: 
simulation FNEC-RT. Open triangles show the gas fraction ($\mg/M_{\rm
tot}$), while stars show the star fraction $\ms / M_{\rm
tot}$. Different colors correspond to different redshifts as detailed
in the legend. The horizontal dashed line corresponds to the baryonic
mass fraction ($\Omega_{\rm b}/\Omega_0 = 0.14$)}
\end{figure*}

In this section, we discuss relations between global properties of 
galaxies in our cosmological simulations, how these depend on the 
detailed physics included in each simulation, and how they compare to 
the observationally established correlations and our analytic 
models of \S \ref{ce}.  

The panels in the figures discussed in this section are arranged in
the following way: left panel - FEC simulation (SN feedback, uniform UV
background, tabulated heating/cooling rates); middle panel - NFEC
simulation (no SN energy feedback, uniform UV background, tabulated 
heating/cooling
rates); right panel - FNEC-RT simulation (SN feedback, 3-D radiative 
transfer). We
do not show the results of the F2NEC-RT simulations in all of the
figures for clarity. In all cases, the results of the F2NEC-RT
simulations are quite similar to those of the FNEC-RT run.

\subsection{Baryonic Content of Simulated Galaxies}

\begin{figure*}[h]
\plotone{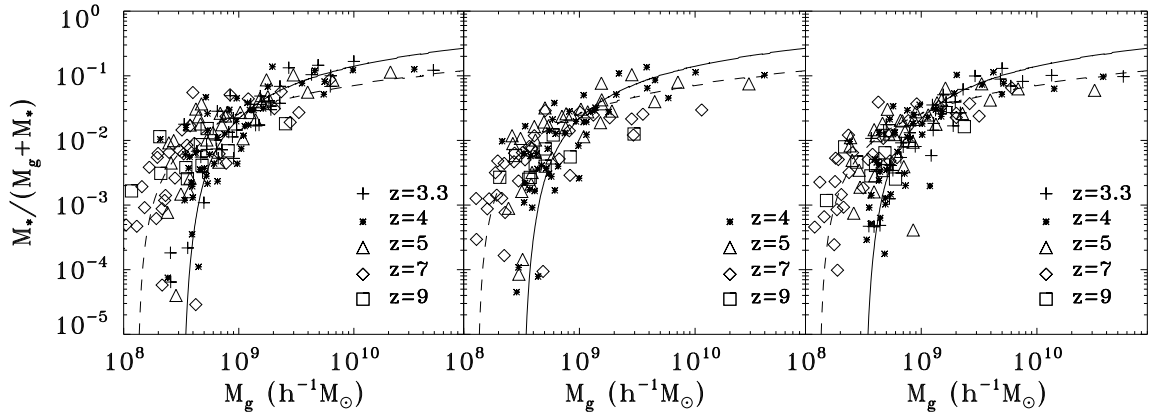}
\caption{\label{MgMst} Star-to-baryonic mass ratio vs. gas mass
for objects in each
of the analyzed simulations. The gas mass here is the total gas in
each halo.  {\it Left panel:\/} simulation FEC;
middle panel: simulation NFEC. {\it Right panel:\/} the FNEC-RT
simulation. Different symbols correspond to different redshifts as
detailed in the legend. The solid and dashed lines show the results of
our analytic model for $z = 4$ and $z=9$,
respectively.}
\end{figure*}

Figure~\ref{MgMtot} shows the gas and stellar mass fractions (total
gas mass over total dynamical mass, and stellar mass mass over total
dynamical mass, respectively) within
the virial radius as a function of the virial mass for all objects in
the simulations.  For systems with $M\gtrsim 10^9\, \msol$, the gas
fractions are approximately independent of mass, while at smaller
masses a variety of effects (e.g., gas stripping, reionization) act to
suppress the gas fractions and increase the scatter.  This result is
qualitatively consistent with the findings of the previous studies
\citep[e.g.,][]{Gn00,CGO01, Tetal03,hoeft_etal06}. Figure~\ref{MgMtot}
also shows that the stellar fractions for the redshifts and masses
under consideration is an order of magnitude smaller than the gas
fractions. This is due both to the inefficiency of star formation
in these small systems and the fact that we analyze them 
at high redshifts, where they did not have the sufficient amount
of time to convert their gas into stars. 
Most interestingly, the stellar fractions drop sharply in systems
with $M\lesssim 5\times 10^9\rm\, M_{\odot}$. The objects with smaller
masses do contain a universal share of the baryons in the form of gas,
but fail to convert a sizeable fraction of it into stars. Note 
that all three simulations shown produced very similar results. 
This demonstrates explicitly that the inefficiency of star formation
at small masses is not due to SN feedback. We note however that
objects in these small-mass systems are resolved with just a few
thousand DM particles. Convergence tests with much higher resolution
simulations are therefore needed to test whether inefficient
star formation at the smallest masses is not a numerical artefact. 

\begin{figure*}[h]
\plotone{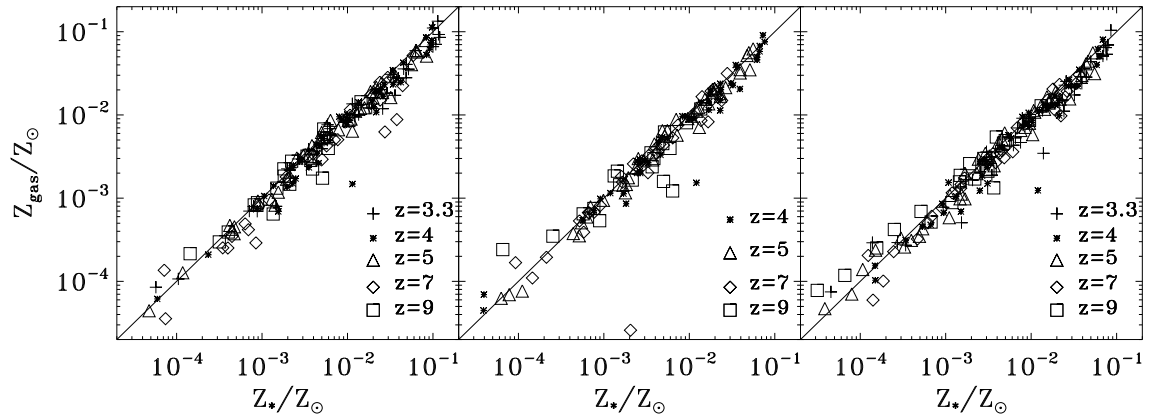}
\caption{\label{ZstZg} Gas metallicity vs. stellar metallicity. 
Left panel: simulation FEC; middle panel: simulation NFEC; right panel: 
FNEC-RT simulation. Different symbols correspond to different redshifts as
detailed in the legend. The solid line corresponds to  $Z_{\rm gas} =
Z_*$. }
\end{figure*}

Figure~\ref{MgMst} shows the star-to-baryonic
 mass ratio of the simulated galaxies
plotted against their gas mass.  A tight correlation between the two
quantities exists, although it cannot be described by a single power
law for the entire range of masses examined here. 
At the high-mass
end, $M_{\rm g} \gtrsim 10^9 \, \msol$, the stellar mass is roughly
proportional to the baryonic mass (and hence, in the gas-rich regime, 
 to the gas mass). 
At lower masses, however, the 
relation exhibits a break and becomes considerably steeper as the
stellar mass of low-mass objects is suppressed \citep[see
also][]{CGO01}.  This correlation is central in identifying the origin
of the scalings between the rest of the global properties we present
below.  The solid and dashed lines represent the prediction of the
model discussed in \S~\ref{ce}, for redshifts $z=4$
and $z=9$, respectively. The agreement with the simulation data is
excellent. Here again we see that the results of the three different
simulations are consistent with each other.

\subsection{Correlations Involving Metallicities}

\begin{figure*}
\plotone{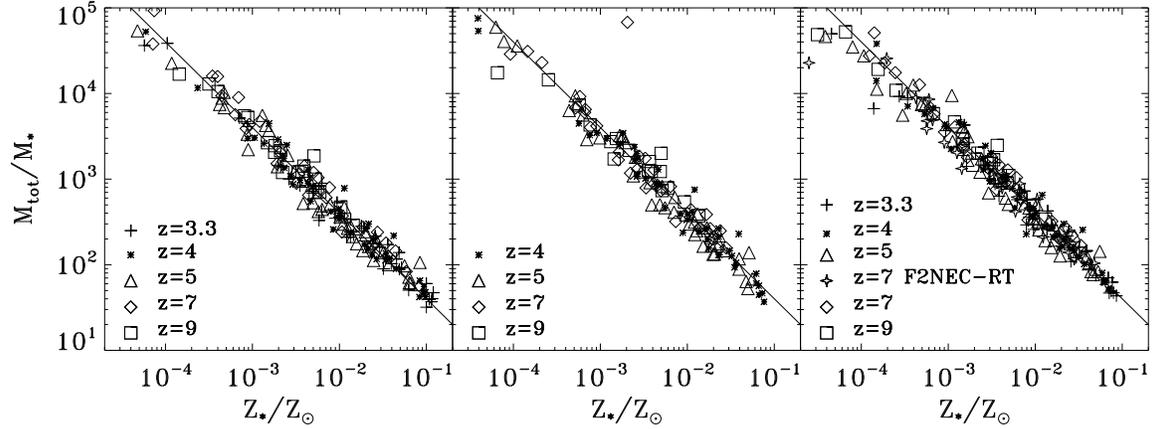}
\caption{\label{MLZ} Ratio between total mass and stellar mass (assumed to be
 proportional to the mass-to-light ratio of each object), plotted
 against stellar metallicity. 
Left panel: simulation FEC; middle panel: simulation NFEC; right panel: 
FNEC-RT simulation. Different symbols correspond to different redshifts as
detailed in the legend. The solid line corresponds to the $M_{\rm
tot}/\ms \propto Z^{-1}$ scaling}. 
\end{figure*}

\begin{figure*}
\plotone{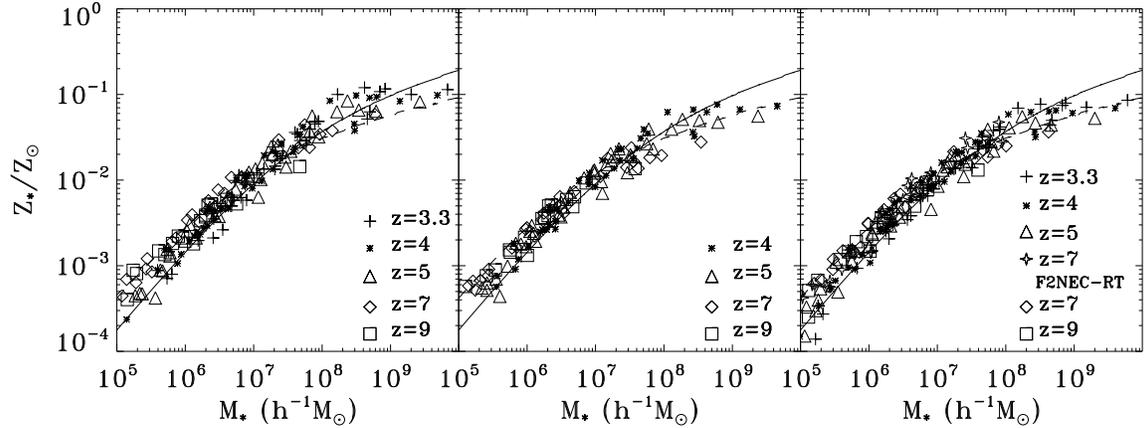}
\caption{\label{ZMst} Stellar metallicity as a function of stellar mass. 
Left panel: simulation FEC; middle panel: simulation NFEC; right panel: 
FNEC-RT simulation. Different symbols correspond to different redshifts as
detailed in the legend.  The solid and dashed lines show the results
of our analytic model for $z = 4$ and $z=9$,
respectively. } 
\end{figure*}

Before we proceed to discuss the correlation of the metallicities in
the simulated galaxies with other observables, we should make a note
about the distinction between stellar and gas metallicities.
Metallicities in stars represent an average metallicity over the star
formation history of the observed object, while gas metallicities are
actually representative of the amount of metals currently present in
the interstellar medium. Observationally, either one or both can be
measured and used.  For example, 
\cite{DW03} use stellar metallicities for dE
galaxies and gas oxygen abundance for dIrr, while \cite{PB02} use
stellar metallicities in all cases.  The metallicity in the gas and in
the stars, although not the same in general, are related to each
other. This is particularly true for the high-redshift galaxies in our
simulations. Figure~\ref{ZstZg} shows the metallicity in the gas
versus the metallicity in stars for every object in our simulations at
different redshifts. The solid line corresponds to the one-to-one
relation $Z_{\rm gas} = Z_*$.  Indeed, for the objects and redshifts
we consider, the metallicity of stars tracks that of the gas 
very closely.

A tight correlation observed for the nearby dwarf galaxies
is the correlation of the dynamical mass-to-light ratio and
metallicity \citep{PB02}. In \S~\ref{mlzmodel} we showed that such
a correlation should naturally arise in low-metallicity dwarf galaxies
without gas outflows and regardless of their star formation rate as
long as their baryon mass is approximately proportional to their total 
mass. Figure~\ref{MLZ} clearly shows that for galaxies in our simulations
the ratio of the total (baryonic + dark matter) to stellar mass is
tightly correlated with metallicity, independently of the details of
star formation and cooling and inclusion of SN feedback. The form of
the correlation, $M/\ms\propto Z^{-1}$ is similar to that
exhibited by the observed dwarf galaxies.  The correlation
arises for objects at the earliest epochs, which
implies that a similar correlation should be expected to hold for 
the observed galaxies at high redshifts.

Basic assumptions in the derivation of this correlation is that $\mg
\propto M$ and $Z\ll Z_\odot$, implying that at the time this 
correlation {\em is being established}, the galaxy is in the gas-rich,
low-metallicity regime. For galaxies in which star formation is
inefficient and which have not lost substantial amount of gas (due to,
e.g., ram or tidal stripping), it is clear that both assumptions are
likely to hold until late times. Such galaxies may move along the
correlation as they evolve, but they will remain on the same line. 
However, such galaxies are not the {\em only} objects which may retain
this correlation until the current cosmic epoch: in galaxies where star
formation is halted early on (when our derivation assumptions are
still valid) and gas is stripped from them, neither the metallicity
nor the dark matter or the stellar mass of the galaxy would be
affected. Hence the correlation between $M/\ms$ and $Z$ would be 
preserved. This would then be a 
typical example of a ``fossil'' correlation, established in the early
phases of the evolution of a galaxy and preserved until the present
cosmic epoch due to lack of additional evolution of the observable
quantities involved. 

Figure~\ref{ZMst} shows the mass-weighted metallicity of stars of the
simulated galaxies as a function of stellar mass. The correlation of
the two quantities is very tight and can be described by a power law
at small masses.  At large stellar masses, $\ms\gtrsim 5\times
10^8\,\msol$, the correlation becomes shallower. This is qualitatively
consistent with the flattening of the $Z-\ms$ relation found in the
SDSS data \citep{Trem04}, although in their observations the break
occurs at much larger masses: $m \sim 3 \times 10^{10}\, \msol$. Note,
however, that this figure should not be directly compared to the
present-day correlation over a wide range of stellar masses because
massive galaxies will undergo substantial evolution in both their
stellar mass and metallicity between $z=3$ and the present, while the
evolution of dwarf galaxies is expected to be much slower. For some of
the dwarf galaxies the evolution can be halted at high redshifts if
their gas is removed by some process (e.g., ram pressure stripping).
Thus, the comparison with the correlations exhibited by the local
galaxies is most meaningful for smaller mass objects.

The solid lines in Figure~\ref{ZMst} show the results of our analytic 
model. The model reproduces the slope of the correlation in
the two mass regimes, the location of the break, and the observed
redshift evolution seen in the simulated galaxies. The correlation in
the model arises mainly due to the inefficiency of star formation in
dwarf galaxies and the existence of star formation threshold, without gas
outflows. Interestingly, direct extrapolation
of the model to $z=0$ indicates that the flattening of the $\ms-Z$
relation shifts to $\ms\sim {\rm few} \times 10^{10}\, \msol$, in
agreement with the SDSS data. However, the reader should keep in mind
that certain assumptions ($Z\ll Z_\odot$,
$\ms \ll \mg$) of our analytic
model are expected to eventually break down at late times (especially
for the higher-mass objects), and hence additional effects may become
important in shaping the present-day properties of higher-mass
galaxies.  

\subsection{Correlations Involving Stellar and Total Masses}

\begin{figure*}
\plotone{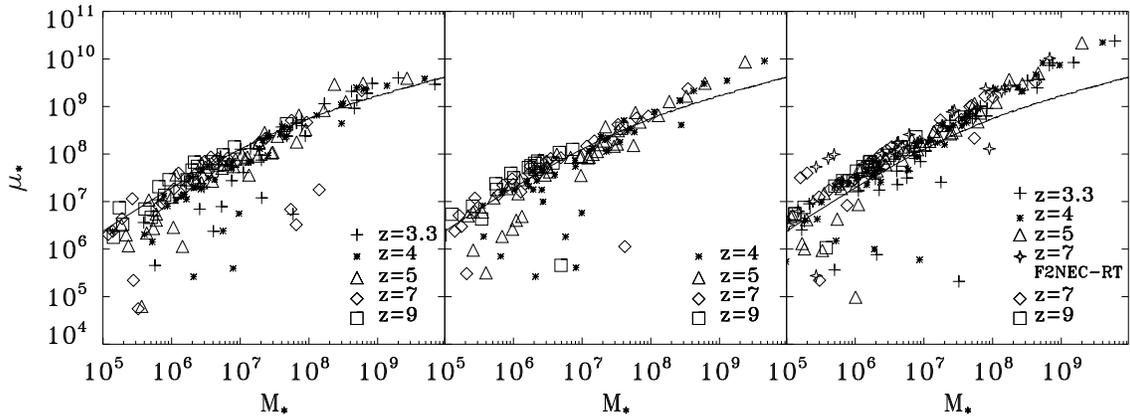}
\caption{\label{SlMst} Stellar mass within $R_*$ over $R_*^2$ (where $R_*$ is the
half-light radius, which includes $50\%$ of the stellar mass of each object),
plotted against stellar mass. Left panel: simulation FEC;
  middle panel: simulation NFEC; right panel: FNEC-RT simulation.
Different symbols correspond to different redshifts as
detailed in the legend. The solid line  corresponds
to the prediction of out analytic model for $z=4$.}
\end{figure*}

Figure~\ref{SlMst} shows the stellar surface density, $\mu_*$, against
galaxy stellar mass. The surface density is calculated from the
simulation data as $\mu_* \propto \ms/R_*^2$, where $R_*$ is the half-light radius, defined as the radius which includes $50\%$ of the
stellar mass.  The data occupy a primary locus in the $\mu_* - \ms$
plane, which is traced by the prediction of our analytic 
model, in which stellar surface density is calculated as
$\mu \propto \ms/r_{\rm th}^2$.  At masses $\ms\lesssim 10^8\, \msol$
there are several outliers, as in this regime star formation becomes
more stochastic. As for the other considered correlations, we 
see that the results of different simulations are consistent, although the NFEC and FNEC-RT simulations do show a higher $\mu_*$ for comparable $\ms$ at the high-$\ms$ end, an indication that in these simulations the stellar populations appear to be more concentrated.  
We should note however that stellar surface densities are probably 
less reliable than global properties, such as stellar mass, because they depend
on the internal structure of the stellar distribution in galaxies
and may be more susceptible to resolution effects.

\begin{figure*}
\plotone{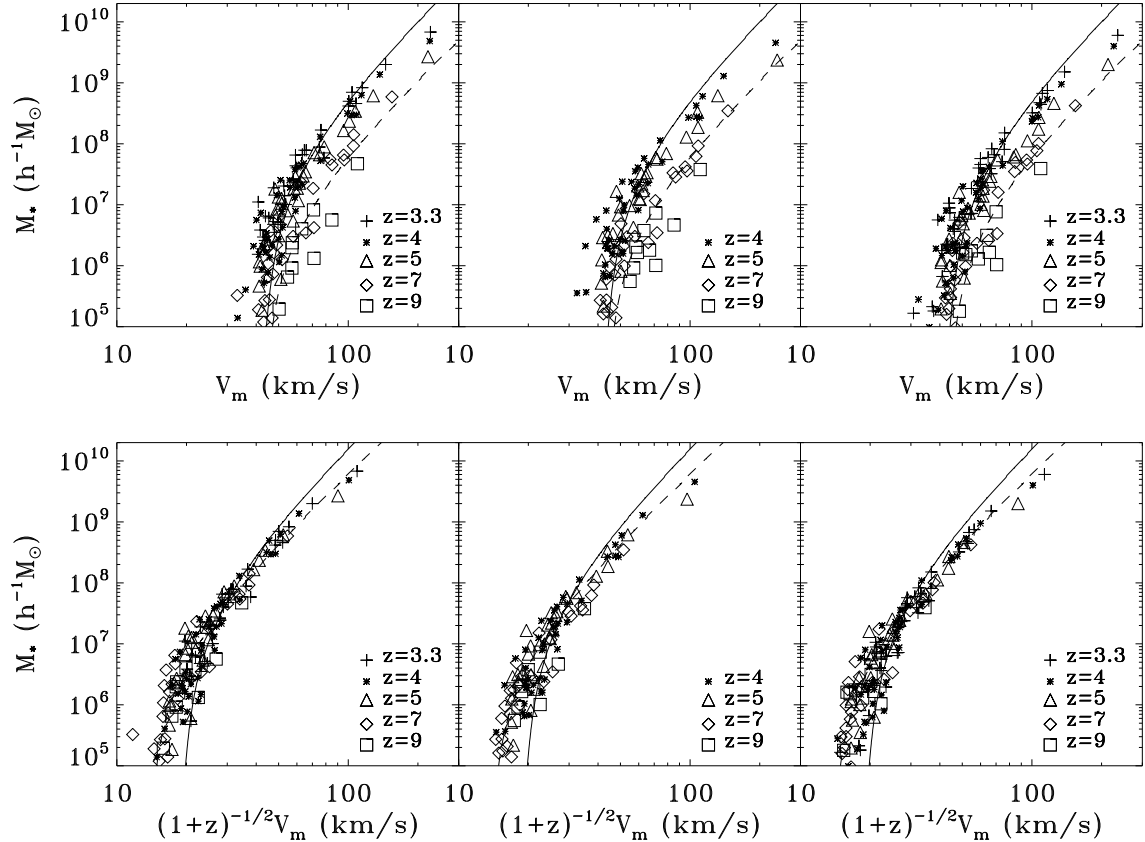}
\caption{\label{MstV} Maximum circular velocity as a function of
  stellar mass (Tully-Fisher relation).  
Left panel: simulation FEC; middle panel:
simulation NFEC; right panel: FNEC-RT simulation.
Different symbols correspond to different redshifts as
detailed in the legend.  The solid and dashed lines show the results
of our analytic model for $z = 4$ and $z=9$,
respectively. In the lower row plots we have shifted $V_{\rm m}$ by a factor of
$(1+z)^{-0.5}$ to correct for the explicit redshift dependence
entering $V_{\rm m}$ through its dependence on the virial radius.
}
\end{figure*}

Figure~\ref{MstV} shows the stellar mass of the objects in our
simulations as a function of their maximum rotational velocity 
(the Tully-Fisher relation). A break of the scaling law
exists at a stellar mass of $\sim 10^8\, \msol$. 
For objects with
stellar masses above the break, the stellar mass scales as 
$V_{\rm m} ^{\alpha}$ with the best fit slopes of
$3.41\pm 0.33$, $3.34\pm 0.33$,
and $3.37 \pm 0.22$ for the FEC, NFEC, and FNEC-RT simulations.
Below the break the $V_{\rm m}-\ms$ relation
steepens considerably. The location of objects on the $\ms - V_{\rm m}$
plane exhibits a considerable shift with redshift. This is 
a trivial consequence of the redshift-dependence of the 
virial density. 

From the virial theorem we have 
\begin{equation}
V_{\rm m} \propto \sqrt{M/R_{\rm vir}} 
\,.
\end{equation}
At the high redshifts under consideration, the universe is 
matter-dominated, and 
the virial overdensity $\delta_{\rm vir} = \rho_{\rm vir}/\rho_{\rm
m}-1 \approx 18\pi^2$ is essentially independent of $z$.
The virial radius is then determined from the relation
\beq
\frac{4}{3}\pi R_{\rm vir}^3 \rho_{\rm m}(z) (1+\delta_{\rm vir}) = M\,.
\eeq 
Given that $\rho_{\rm m}(z) = \rho_{\rm m,0}(1+z)^3$, we get 
\beq
R_{\rm vir} \propto \frac{M^{1/3}}{1+z}\,.
\eeq
and hence
\begin{equation}
V_{\rm m} \propto M^{1/3}\sqrt{1+z}\,.
\end{equation}

In the lower panels of Figure~\ref{MstV}, we have scaled the $V_{\rm
 m}$ values of all data points of the upper panel by $\sqrt{1+z}$~to
 correct for this redshift evolution.  Using $M \propto \mg$ and the
 $\ms=\ms(\mg)$ dependence calculated using our analytic model, we get
 the solid and dashed lines shown in Figure~\ref{MstV}, for $z=4$
 and $z=9$, respectively. Once again, the model describes the results
 of the simulation very well.

\begin{figure*}
\plotone{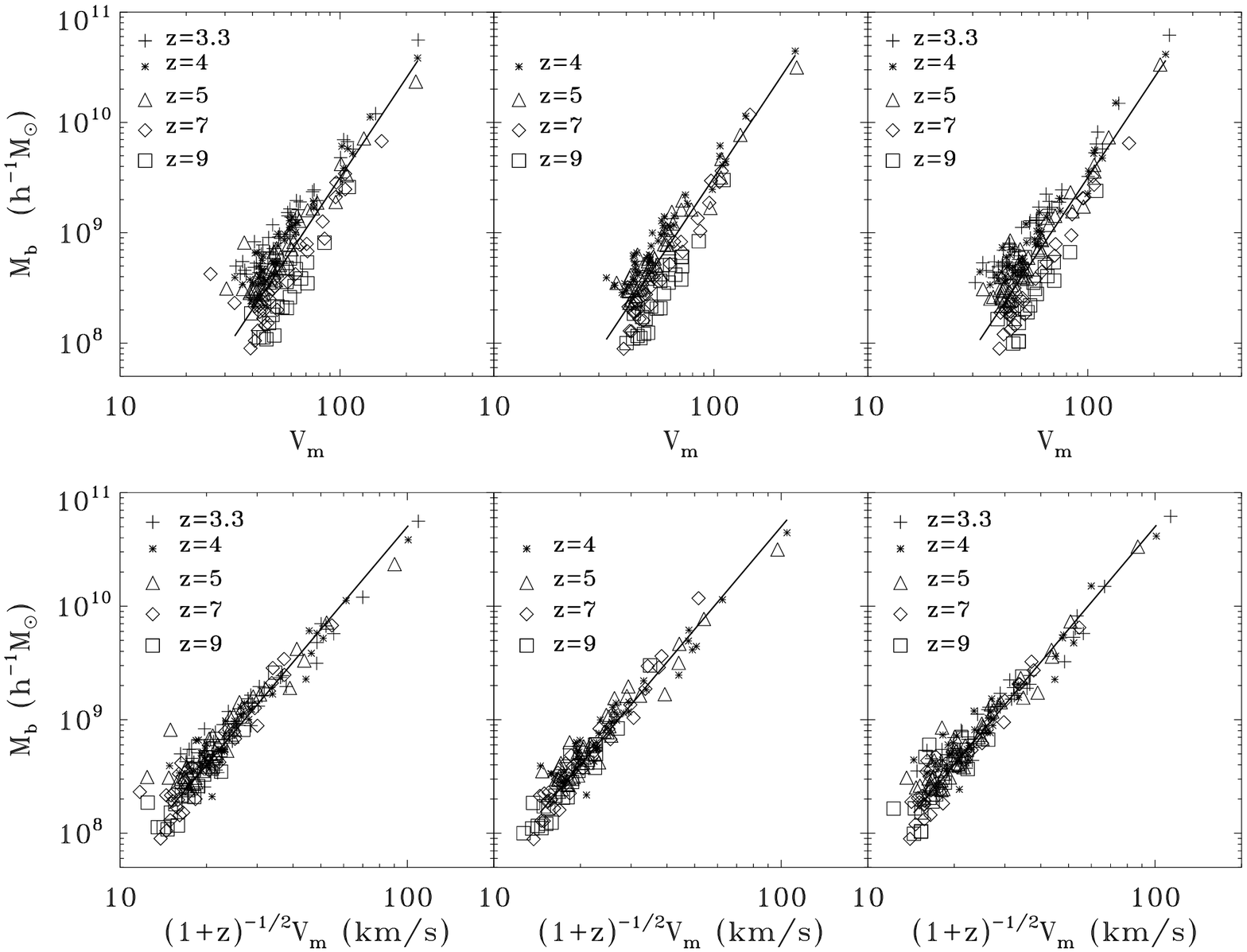}
\caption{\label{MbV} Maximum circular velocity as a function of
  baryonic mass ({\em baryonic} Tully-Fisher relation). 
Left panel: simulation FEC; middle panel: simulation NFEC;
  right panel: FNEC-RT simulation. Different symbols correspond to
  different redshifts as detailed in the legend. The solid line
  corresponds to the $M_{\rm baryon} \propto V_{\rm m}^3$ scaling.  
In the lower row plots we have shifted $V_{\rm m}$ by a factor of
$(1+z)^{-0.5}$ to correct for the explicit redshift dependence
entering $V_{\rm m}$ through its dependence on the virial radius. }
\end{figure*}

Finally, Figure~\ref{MbV} shows the  {\em baryonic} Tully-Fisher
relation, where the total baryonic mass (stars and gas) 
of each object is plotted against $V_{\rm m}$. In this case,
the $M_{\rm baryon} \propto V_{\rm m}^3$ scaling (represented by the 
solid line) shows no evidence of a break. This is similar
to the behavior of the observed dwarfs, which exhibit a break 
in the $\ms-V_{\rm m}$ relation, but a tight power law $M_{\rm b}-V_{\rm m}$
relation (see \S~\ref{obs}). 

\subsection{The Effective Yield}\label{syeff}

\begin{figure*}
\plotone{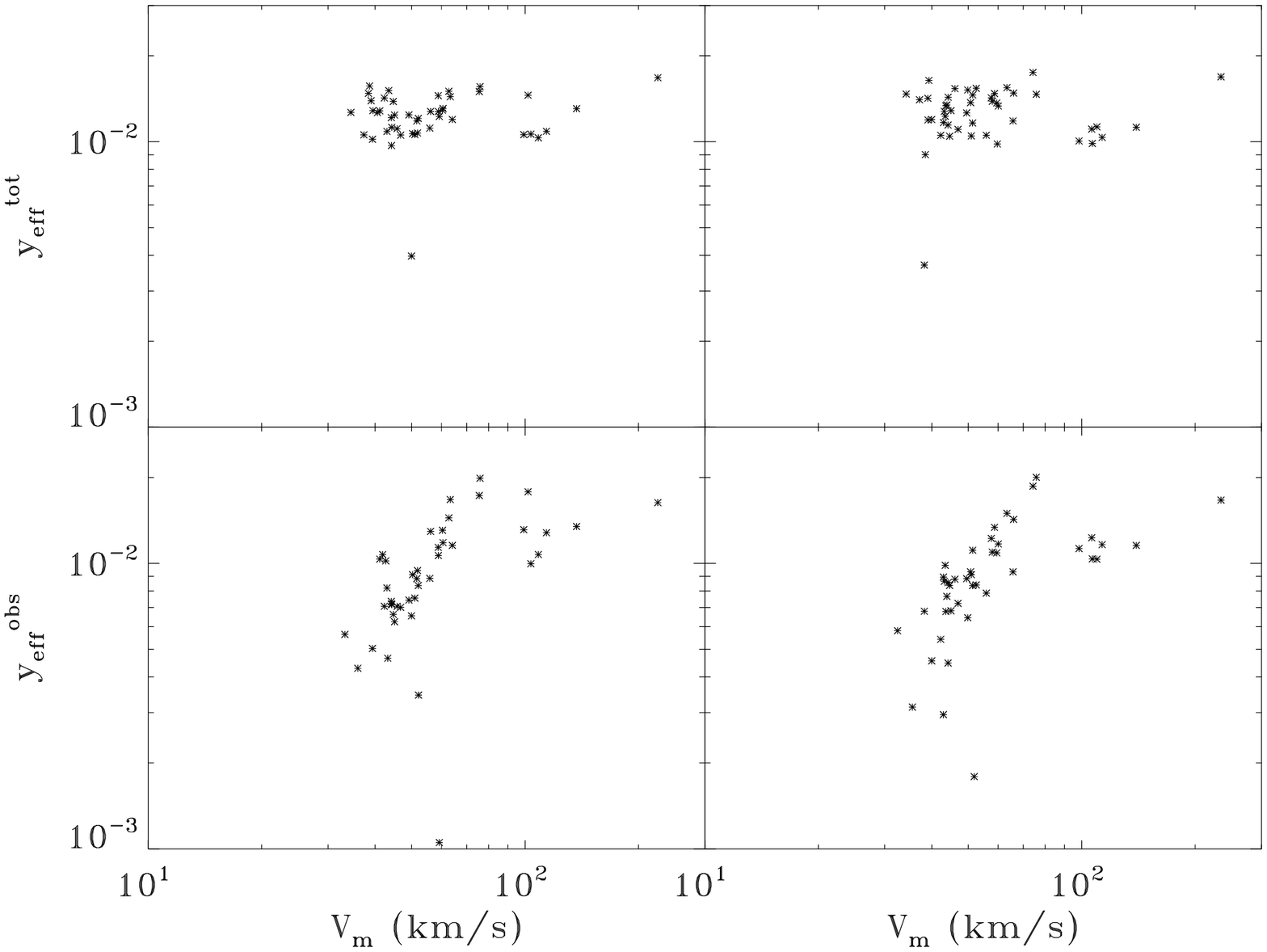}
\caption{\label{fig:yeff} Effective yield as a function of circular 
velocity of galaxies in our simulations at $z=4$. Left column: simulation FEC 
(with supernova energy feedback); right column: simulation NFEC 
(no supernova energy feedback). Upper panel: effective yield 
$y_{\rm eff}^{\rm tot}$ calculated taking into account the mass of gas, stars, 
and metals within the virial radius of each object; lower panel: 
effective yield $y_{\rm eff}^{\rm obs}$ calculated using the metallicity of
the stellar disk and the gas fraction calculated using all cold gas
within the halo.}
\end{figure*}

The effective yield, defined as
\begin{equation}\label{yef}
y_{\rm eff} = \frac{Z}{\ln (1/f_{\rm g})},
\end{equation}
has been widely used as a diagnostic of the evolution of the baryonic
component of galaxies, and more specifically as a ``litmus test'' of 
the validity of the closed-box approximation.  Here, $f_{\rm
g}=\mg/(\ms+\mg)$ is the fraction of baryons in the gas phase. Under the
closed-box assumption, the effective yield is always equal to the true
yield $y_{\rm true}$, defined as the mass in newly synthesized metals
returned to the ISM by a stellar population normalized to 
the stellar mass of this population locked-up in stellar remnants and
long-lived stars.\footnote{The true yield is related to the mass of
newly synthesized metals normalized to the {\em total} mass of the
parent generation of stars, $q_Z$: $y_{\rm true} = q_z/a$, where $a$
is the lock-up fraction.}

\citet{Edm90} showed that $y_{\rm eff}$ cannot exceed $y_{\rm true}$.
Values of $y_{\rm eff}$ lower than $y_{\rm true}$ are indicative of
the deviations from the closed-box model due to either inflows or
outflows of gas and metals.  Observationally, dwarf galaxies ($V_{\rm
m} \lesssim 100 {\rm \, km/s}$) have low values of $y_{\rm eff}$. This
has been interpreted as strong evidence for significant outflows of
metals and/or gas, which can reduce the metallicity and/or the gas 
fraction and are expected to be more prevalent in smaller
objects with shallower gravitational potentials
\citep{Garn02,Trem04,Pil04}.
 
In the previous section we showed that global correlations of galaxy
properties arise in our simulations without SN-driven outflows.  We
have shown that these correlations can be reproduced with an open-box
evolution with cosmological mass inflow but no outflows and 1)
mass-dependent gas distribution given by equation~(\ref{rd}) and 2)
star formation obeying the Kennicutt law {\em with a surface density
threshold}.  In this model the correlations, similar to those of
observed galaxies, arise due to the increasing fraction of gas at
surface densities below the star formation threshold density (and,
correspondingly, inefficiency of star formation) in smaller mass
objects. This trend, in turn, is due to inefficiency of cooling and
more extended gas distributions for smaller masses. 
Given that the observed values of the effective yield are considered
to be evidence for outflows, in this section we compare the predictions of 
our simulations for the values of the effective yield. 

Upper panels of Figure~\ref{fig:yeff} show the effective yield as a
function of circular velocity for the objects in simulations FEC (with
feedback) and NFEC (no feedback) at $z=4$. Here $y_{\rm eff}^{\rm tot}$ is
calculated taking into account all of the gas and metals inside the
virial radius of each object. As expected, in the absence of outflows
the effective yield is constant and is close to the true yield adopted
in simulations.

However, we should recognize that in observations 
the gas fraction only includes gas of densities and
temperatures accessible through HI observations \citep[e.g.,][]{Trem04,Garn02,Pil04,lee06}.  If a fraction of metals resides in the gas phase that is not observed, the measured effective yield will be lower, just as in the case when the enriched gas is ejected from the system completely. 
The lower panels in
Figure~\ref{fig:yeff} show the values of $y_{\rm eff}^{\rm obs}$ calculated
using metals only within the stellar extent, defined as the
radius that includes $90\%$ of the total stellar mass, and cold gas
only ($T\lesssim 10^4$K).  
The metallicity entering the ``observable'' 
effective yield is calculated as the ratio of the metal 
mass within the $90\%$ stellar mass radius, and the gas mass 
within the same radius. The gas entering the gas fraction is all gas
within the halo with $T\lesssim 10^4$K.
In this case,
there is a marked decrease of $y_{\rm eff}^{\rm obs}$ with decreasing
circular velocity at $V_{\rm m}\lesssim 100\rm\,km\,s^{-1}$. The 
values of the effective yield for the small-mass systems are now
comparable to those measured for dwarf galaxies. 
The decreasing effective yield for lower mass systems suggests
that a fraction of metals in these dwarf galaxies is not in the
observed cold gas. Given that the effective yield within the virial
radius is constant and is equal to its true value, the extra 
metals must reside within the virial radius but in a hot phase gas,
detectable perhaps only by its absorption in spectra of background
sources and which is thus not included in the calculation of the gas fraction.

In order to reproduce the
correct trend with galaxy mass, low-mass objects should have a larger
fraction of the metals residing in the unobservable gas phase. 
\begin{figure*}
\plotone{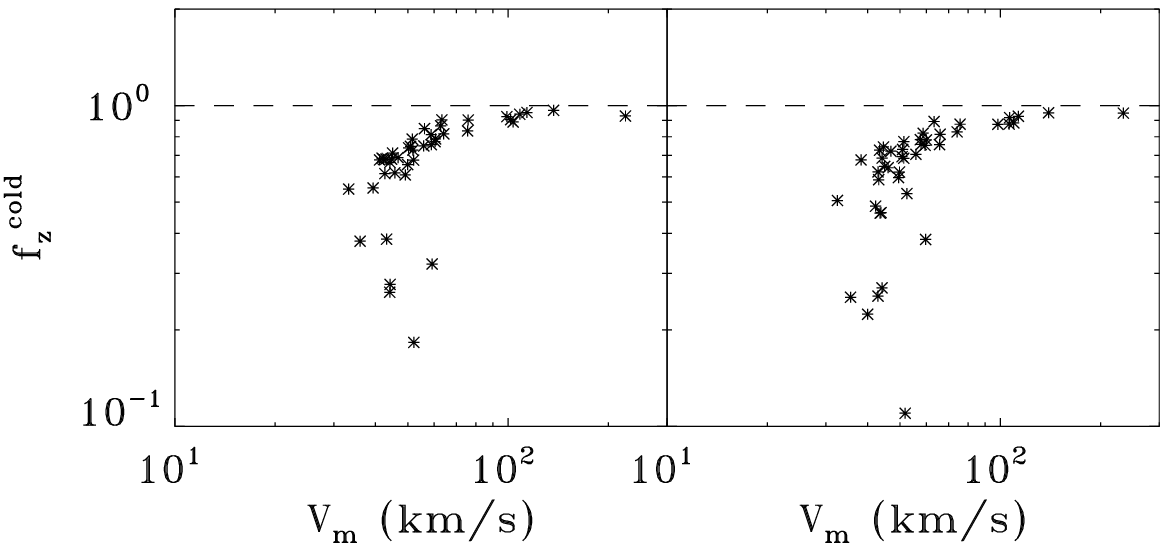}
\caption{\label{fig:yeffratios} 
Fraction of metal mass within the cold gas disk.  
Left column: simulation FEC 
(with supernova energy feedback); right column: simulation NFEC 
(no supernova energy feedback). }
\end{figure*}
Figure \ref{fig:yeffratios} shows 
the fraction of the metal mass within the cold gas
disk, $f_Z^{\rm cold}$  and shows that our
simulated galaxies do indeed exhibit such a trend. For higher-mass
galaxies, this fraction is close to unity, indicating that most of the
metals are in the cold gas phase.  For low-mass galaxies, however,
this fraction is significantly below unity, which means that a
significant fraction of metals is in the hotter, unobservable gas. It
is this trend that is responsible for the decreasing effective yield
values for smaller mass galaxies in the bottom panels of
Figure~\ref{fig:yeff}. The left and right columns, corresponding to
simulations FEC and NFEC respectively, show similar trends, which
shows that supernova-driven outflows are not   
responsible in our simulations, and therefore are not necessarily 
responsible in nature, 
for the presence of metals outside of the cold phase.

There is widespread turbulence in the interstellar medium of our
simulated galaxies, driven
by gravitational instabilities, cold accretion flows
\citep{birnboim_dekel03,keres_etal05} reaching well inside the virial
radius and stirring up the gas. However, the dominant source of both
turbulence in the cold gas as well as mixing of metals in the hot
phase are most likely mergers with other galaxies and dark matter
subhalos, frequent at high redshifts. There is also the possibility
that numerical diffusion also contributes to some extent to the
efficient spread of metals within the virial radius of each
object. Higher resolution simulations are required to quantify this 
effect.

If this picture is also applicable in nature, the prediction of our
model would be that, in the case of small dwarfs with 
low values of the observed effective yield,  
the gas well outside of the extent of the observable cold disk
should be
significantly enriched with heavy elements.  
Although to our knowledge
there are no current observational constraints on the metallicity of
such gas\footnote{Some existing measurements of the metallicity of
neutral HI gas for dwarf galaxy I Zw 18
\citep[e.g.,][]{aloisi_etal03,lecavelier_etal04} indicate that the
metallicity of the neutral gas is comparable or somewhat below that of
the HII gas for which the metallicities are typically measured. These
measurements of the neutral phase metallicity, however, are still
limited to regions within the stellar extent. It is not yet clear
whether results for this 
very low metallicity galaxy are also true for dwarf galaxies in
general.}, metallicity 
measurements in absorption and tests of our conjecture may be possible
in the future.  The current measurements for $\sim L_{\ast}$ galaxies
indicate that cold gas is significantly enriched to large radii where
little or none in situ star formation should be occuring
\citep{chen_etal05}.  In addition, direct measurements of metal
abundances at different radii show that metallicity gradients in
dwarfs galaxies are weak to at least four exponential disk scale
lengths \citep[][and references therein]{lee_etal06b}. This may 
indicate mixing of metals in the gaseous disks is indeed
efficient.

\section{Discussion}\label{disc}

Before a detailed comparison with observational results is attempted, it is
important to consider how the results that we obtain for simulated 
galaxies at high-redshifts can relate to the galaxies observed at lower
redshifts. Our analytic model shows that the correlations we observe
in the simulated galaxies arise because galaxies are gas rich and 
star formation is increasingly inefficient for lower mass systems (which
is due to the existence of density threshold for star formation). 
These results thus are not applicable to galaxies which formed 
significant fraction of their stars under different conditions. 

However, we argue that our results are applicable to certain types of
the nearby dwarf galaxies. All of such galaxies could have undergone
outflow-free
evolution in the gas-rich regime at high redshift, similar
to the our simulated galaxies. Their subsequent fates could be roughly
classified into three scenario. Some galaxies could have continued to
grow their mass and form stars and consumed most of their gas. Such
galaxies could have formed most of their stars in the regime where our
results are not applicable. This scenario likely applies for those
small-mass, high-$z$ systems that have experienced significant mass
growth and became massive objects today.  Another scenario is that
star formation could have been truncated at high redshifts, while
still in the gas-rich regime, by ram pressure and tidal forces from
the nearby massive galaxy. Such scenario is probably applicable for
the nearby dwarf spheroidal galaxies
\citep{KGK04,mayer_etal07}.  Finally, the third scenario is
that the mass of a high-$z$ dwarf system remained low throughout its
evolution. In this case most of the stars could have formed in the
low-efficient mode of star formation and the galaxy may have remained
in the gas-rich regime until $z=0$.

In either one of the latter two scenarios, high-redshift correlations
between metallicity, total mass, stellar mass and surface brightness
would be preserved to $z=0$. Note that both nearby gas-poor dwarf
spheroidals and gas-rich dwarf irregulars exhibit similar correlations
of their global properties such as luminosity and metallicity
\citep[e.g.,][]{GGH03}.  The slopes of these correlations are indeed
very close to the ones we find in our models of high-redshift
objects. This may indicate that the evolution of these galaxies has
proceeded according to the last two scenarios.

\begin{figure*}
\plotone{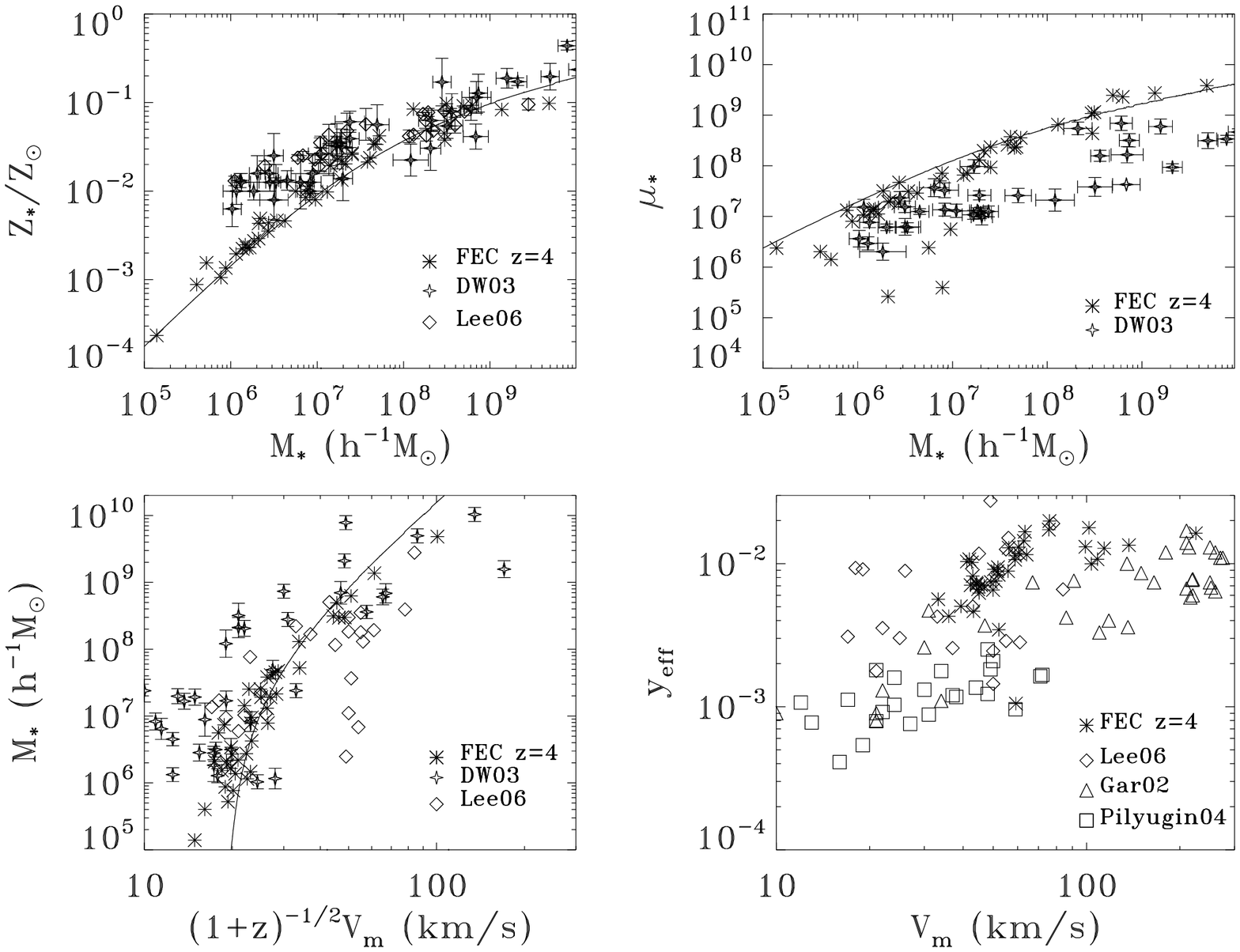}
\caption{\label{Obs} Observational data overplotted with results from 
the FEC simulation (z=4). The solid line corresponds to the analytic
model prediction, also for $z=4$. 
Upper left panel: metallicity vs stellar mass. 
Observational data from \cite{DW03} and \cite{lee06}.
Upper right panel: surface brightness vs stellar mass. Observational
data from \cite{DW03}. Lower left panel: Stellar mass vs
maximum rotational velocity (Tully-Fisher
relation) corrected for explicit redshift evolution. Observational data
from \cite{DW03} and \cite{lee06}. Lower right panel:
effective yield vs maximum rotational velocity. Observational data
from \cite{lee06}, \cite{Garn02} and \cite{Pil04}.}
\end{figure*}

It is thus interesting to examine how our simulation results
high-redshift results compare with observations of dwarf galaxies in
the present-day universe. In Figure~\ref{Obs} we plot observed
properties of dwarf galaxies, the $z=4$ results from the FEC
simulation, and the predictions of the analytic model for the same redshift. The upper left
panel shows the metallicity as a function of stellar mass.
Observational data are taken from \cite{DW03} (in turn taken from
\citealp{mateo98}, \citealp{vdb00} and references therein) and from
\citep[][their Table 1]{lee06}.  In the case of the \cite{lee06} data,
we used the conversion relation fitted by \citep[][their
Eq.~16]{simon_etal06} to convert $12+\log({\rm O/H})$ to $[{\rm
Fe/H}]$. The upper right panel shows the surface brightness as a
function of stellar mass.  In the lower left panel, we plot the
stellar mass as a function of the maximum circular velocity (the
Tully-Fisher relation), corrected against explicit evolution with
redshift. Finally, in the lower right panel we show the effective
yield as a function of maximum circular velocity. The overplotted
observational points are taken from \citep[][their Table 5]{lee06},
\citep[][their Table 4]{Garn02} and \citep[][we have used data on
oxygen abundances and gas fractions from their Table 7 to compute the
effective yield]{Pil04}. The effective yield in the simulated objects
has been calculated using the metals and gas within the stellar disk
to calculate the metallicity, but all of the cold ($\lesssim 10^4 {\,
\rm K}$), HI-observable gas in each halo to calculate $f_{\rm gas}$,
so as to be directly comparable with the observational data.  Although
the scatter is appreciable both within the same dataset but most
prominently between different datasets\footnote{\cite{Pil04} and
\cite{Garn02} have a large fraction of their samples overlapping,
often deriving very different metallicities for individual
cases. Thus, the scatter between different datasets in the case of the
effective yield is representative of systematic uncertainties in such
determinations.}, the correlations present in the simulation results
are generally very similar with the ones found in the observational
data.

We have shown in \S \ref{res} 
that in all of our models, 
tight correlations, similar to those observed in present-day dwarfs, 
are developed already at very high redshifts between
observables of low-mass galaxies, and they 
are preserved at least until $z \sim 3$. 
Energy feedback for supernovae and galactic winds 
was not found to be either necessary or important for 
establishing these correlations. 
Our interpretation of these correlations in our models and the corresponding 
ones in observed systems is that they occur as a result of the 
inefficiency of star formation in low-mass systems, due to inefficient
cooling and the associated low gas surface-densities
(a concept consistent with observations suggesting that 
in low-mass galaxies the surface density of
cold gas is typically low, e.g., 
\citealp{vanzee_etal97,martin_kennicutt01,auld_etal06}).
As a result, stellar mass and
metal enrichment are increasingly suppressed at lower 
masses, resulting in the observed correlations between stellar mass
and other observables. 
The inefficiency of star formation below the critical
gas surface density also helps to explain the observed gas fractions
as a function of galaxy luminosity and surface brightness
\citep{vdb00}, the stellar truncation radii and gas extent in the
observed disks \citep{vdb01}, the shallow slope of the faint-end of
the galaxy luminosity function \citep{verde_etal02,croton_etal06},
the paucity of faint dwarf galaxies in the Local Group
\citep{verde_etal02,KGK04}, and observations of extreme HI-dominated dwarfs 
\citep{begum_etal05,begum_etal06,warren_etal04,warren_etal06}, which  
are not deficient in baryons but are nevertheless underluminous. 
These results imply that the star
formation law governing conversion of cold gas into stars, rather than 
SN-driven outflows, is the dominant factor in shaping properties of
faint galaxies. 

In addition, there are observational indications that feedback 
does not in fact drive significant amounts of gas out of low-mass galaxies. 
Given that lower-mass galaxies would be more susceptible to strong
supernova-driven gas outflows, one could expect that gas fractions
should decrease with decreasing stellar mass. However, the opposite
trend is observed \citep{Mg97, BDJ00,Garn02,geha_etal06}. In fact, 
many of the smallest galaxies have gas fractions in excess of 0.8-0.9
\citep{geha_etal06,lee06}.

At the same time however, there is indirect evidence for the importance 
of metal loss
in winds. The high metallicity of diffuse intergalactic gas in groups
and clusters \citep[e.g.,][]{renzini_etal93} and in the intergalactic
medium around galaxies \citep[e.g.,][]{adelberger_etal05} is generally 
attributed to the action of winds. The winds may also play a role 
in faster evolution of the stellar mass-metallicity relation \citep{derossi_etal06}.

An interesting additional constraint comes from the observed HI mass
function (HIMF) of galaxies \citep{zwaan_etal05}. Its small-mass end
is significantly shallower than the expected mass function of dark
matter halos, implying that some sort of suppression of the gas content in
halos is needed. At this point, there is no accepted explanation of the
shallow small-mass end of the HIMF, although \citet{mo_etal05} have
recently argued that it can be explained by a combination of gas
heating by the cosmic UV background and preheating of accreting gas by
filaments and pancakes in which galaxies are embedded. It is
interesting that regardless of the nature of suppression mechanism it
should be almost independent of halo mass. Strong mass dependence of
gas suppression would likely introduce a break in the baryonic Tully-Fisher
relation, which is not observed.

From a chemical evolution point of view, outflows would constitute 
a departure from closed-box evolution. 
The quantity that has been traditionally used to quantify the metallicity
evolution of a galaxy and its deviation from the closed-box model is the 
effective yield, $y_{\rm eff}$ \citep{Garn02, Pil04, Trem04}. 
Lower-mass galaxies ($V_{\rm m} \lesssim 100 {\rm \, km/s}$) have in fact 
been observed to have an effective yield lower than the closed-box model 
prediction for their observed gas fraction, and this has been traditionally 
interpreted as an indication for metal outflows. Recently,
\cite{Dalc06} 
showed that outflows of gas with the same
average metallicity as the host galaxy can only have a minor effect on
the effective yield and cannot reproduce the observed reduction of
$y_{\rm eff}$ with respect to the closed-box value in low-mass
objects, and suggested that metal-enhanced outflows may instead be
responsible for this effect.  

In this paper, we show that there is an alternative explanation
for the trend for decreased effective yields at the lowest-mass
objects.  The effective mixing in the ISM at these high redshifts
spreads the metals produced through star formation to larger regions
in small galaxies than are accessible through observations. When the
effective yield in our simulated galaxies is calculated taking into
account gas, stars and metals out to the virial radius of each object,
no dependence of the effective yield with mass is observed.
Conversely, when we calculate the effective yield taking
into account only the gas, stars and metals that would be accessible to
observations, 
a trend of decreasing effective yield with mass is revealed,
similar to the one seen in observational data. This trend becomes much
less pronounced in higher-mass galaxies, consistent with the findings
of \cite{Erb06} who found no appreciable dependence of effective
yield on stellar mass for objects with $\ms \gtrsim 10^9 \msol$ and for
$z \sim 2$. 

We would like to stress again that gas outflows escaping the host halo
are not required to explain this trend, although they may affect the
exact form of the relation (normalization and slope) and its
evolution. Note however that gas flows {\em within} the host halo may
be important in spreading the produced metals to regions considerably
more extended than the stellar disk.  The consistency between results
from our simulations with and without energy feedback from supernovae
suggests that some other process drives turbulence and
large-scale motions of the interstellar medium, such as gravitational
instabilities, galaxy mergers, or cold accretion flows.  This
explanation of the trend of effective yield with galaxy mass can be
potentially tested by measurements of the metallicity of the gas in
the outskirts of the gaseous disks of dwarf galaxies.

\section{Conclusions}\label{sum}

In this paper, we use a suite of high-resolution cosmological
simulations to examine the role of different physical processes in
establishing the observed scaling relations of dwarf galaxies. All
simulations included a recipe for star formation and metal enrichment
of the ISM. Three simulations (FEC, FNEC-RT, F2NEC-RT) additionally
included energy feedback from supernovae to the ISM. Simulations FNEC-RT and F2NEC-RT 
followed detailed 3-D radiative transfer from individual UV sources
and the chemistry network of ionic species of hydrogen, helium, and
molecular hydrogen to determine the radiative heating and cooling of
the baryonic gas, while the other two simulations used tabulated
heating and cooling rates and a uniform ionizing background.  In order
to interpret our simulation results, we develop and use an analytic
model, which follows the evolution of the dynamical
mass, gas, and stellar components as well as the metallicity of
galaxies. We assumed an open box model with no outflows but with
accretion of mass at a rate proportional to the galaxy mass, and star
formation following the Kennicutt law with a critical density
threshold for star formation.  Our main results and conclusions can be
summarized as follows.
\begin{itemize} 

\item In all simulations, correlations between global
quantities, such as the stellar mass-metallicity relation, 
very similar to those observed for the nearby dwarf
galaxies arise in simulated galaxies with stellar masses 
$10^5 \msol \lesssim \ms \lesssim 10^8 \msol$ for $z \gtrsim 3.3$. 

\item In galaxies of higher masses these correlations exhibit a break
      and the dependence of global properties on stellar mass flattens
      off. A similar flattening of the correlations is also found in
      observational data, although the location of the break may have
      evolved to higher masses by the present epoch
      (e.g., \citealp{Trem04}).  

\item The key finding of this study is that neither the inclusion of
supernova energy feedback, nor the inclusion of 3-D radiative transfer
significantly affects any of these correlations, a strong indication
that thermal and radiative processes (as well as associated galactic
outflows and winds) are not required for the observed
correlations to arise.  Using our analytic model, we show that these
correlations in the simulated galaxies arise due the increasing
inefficiency of star formation with decreasing mass of the object.
The inefficiency of conversion of gas into stars and its dependence on
mass are due to the critical gas column density threshold of
the star formation law.

\item We thus argue that trends similar to those exhibited by observed
  dwarf galaxies can arise without loss of gas and metals in winds,
  although winds may affect the exact form of the relation
  (normalization and slope) and its evolution.

\item The observed trends of a decreasing effective yield with
decreasing galaxy mass can be
reproduced in our simulated galaxies reasonably well, and without
winds, if the effective yield is calculated taking into account only
the gas, stars, and metals that would be accessible to observations.
We show that this trend is due 
to efficient mixing in the ISM, driven by disk instabilities, mergers, 
and possibly cold accretion flows. The efficient mixing redistributes
the metals far beyond the stellar extent in smaller mass galaxies. 
Thus a significant
fraction of metals in small galaxies 
is outside the stellar extent leading to a decrease of the estimated
yield if only metals within the stellar radius are taken into account. 
This explanation can potentially be tested by measurements of the
gas metallicity in the outskirts of the gaseous disks of dwarf galaxies. 
\end{itemize}

\acknowledgements 

We would like to thank referee of this paper for detailed constructive
criticisms that led to significant improvements of the presentation. 
We are very grateful Marla Geha for useful discussions and for
sharing results prior to publications, Andrea Macci\`o for providing
us with corrected values of $V_{\rm m}$ for Local Group dwarfs, and
Vasiliki Pavlidou for helpful discussions. We are grateful to Julianne
Dalcanton, Josh Simon, and Francisco Prada for stimulating discussions
of their results and Simon White, Volker Springel, Licia Verde,
Hsiao-Wen Chen, Justin Read and Greg Bryan for helpful comments on the draft of
this manuscript. AVK and NYG are grateful to the Institute for
Advanced Study and Aspen Center for Physics for hospitality during the
completion of this paper. This work was supported in part by the DOE
and the NASA grant NAG 5-10842 at Fermilab, by the HST Theory grant
HST-AR-10283.01, by the NSF grants AST-0206216, AST-0239759, and
AST-0507666, and by the Kavli Institute for Cosmological Physics at
the University of Chicago.  Supercomputer simulations were run on the
IBM P690 array at the National Center for Supercomputing Applications
(under grant AST-020018N) This work made extensive use of the NASA
Astrophysics Data System and {\tt arXiv.org} preprint server.

\appendix
\section{Gas Mass Accretion Rate}\label{gma}

In the spherical collapse model, 
an overdensity of mass $M$ at a time $t$ accretes matter at a
rate $F = dM/dt \propto M$
\citep{fillmore_goldreich84,Ber85}.\footnote{Although analytic
solutions were derived for an Einstein - de Sitter universe, the
result holds under spherical collapse in any cosmological model. }
 Such a relation is also consistent with the results of 
cosmological simulations, in which halo mass grows with 
time as $M (z) \propto \exp[-Cz]$ \citep{W02}, 
which gives $dM/dt=f(z)M$.
In general, $f(z)$ depends nontrivially on cosmic time 
(with the exception of the Einstein-de
Sitter universe where the evolution is self-similar and the 
$f$ scales with time simply as $\propto t^{-1}$), as well as the cosmological
parameters. 

 Since we consider early epochs in the evolution of galaxies
we assume that the baryon mass is dominated by gas. 
For example, for the simulated halos in the redshifts of
interest: $\mg \gtrsim 10 \ms$. We assume thus that 
$\mg \approx M_{\rm b} = \Omega_{\rm b}M/\Omega_{\rm m}$, so 
\beq\label{acrate}
F = f(z) \mg\,.
\eeq
An expression for $f(z)$ is obtained by
using the following simple recipe:
\begin{equation}\label{simrec}
F = 4 \pi [r_{\rm vir}(z)]^2 (1+\delta_{\rm ext}) \rho_{{\rm m},z}(z) 
v_{\rm acc}(z)\,,
\end{equation}
where $\rho_{{\rm m},z} = \rho_{{\rm m},0}(1+z)^{3}$ 
is the mean matter density of the universe at
redshift $z$, $r_{\rm vir}(z) = [3M/4\pi\rho_{{\rm m},z}(1+\delta_{\rm
vir})]^{1/3}$ is the virial radius of an object of mass $M$ at redshift $z$,
$\delta_{\rm vir}$ is the virial overdensity (equal to $18 \pi
^2\approx 178$ for the high redshifts considered here), $1+\delta_{\rm
ext}$ is the local compression factor right outside the virial radius
of the object, and $v_{\rm acc}(z)= \sqrt{2GM/r_{\rm vir}(z)}$
is the free-fall velocity\footnote{Note that $r_{\rm vir}\propto
M^{1/3}$ and $v_{\rm acc} \propto M^{1/2}r_{\rm vir}^{-1/2} \propto
M^{1/3}$ so $F \propto r_{\rm vir}^2 v_{\rm acc} \propto
M$.}. Equation (\ref{simrec}) is only approximate, and we use the
value of $(1+\delta_{\rm ext})$ to set its exact normalization so that
it adequately reproduces our simulations results, 
which we find to be the case for
$1+\delta_{\rm ext} \approx 20$. Similar
values of the external compression factor used in analytic
cosmological accretion models yield results in general agreement with
the accretion rates and energetics of objects in cosmological
simulations \citep{PF06}.

\section{Star Formation Rate}\label{sfmg}

To calculate $\psi(\mg)$ we assume that gas in each halo settles into
an exponential disk obeying the empirical Kennicutt star formation
law\footnote{Although star formation in real galaxies may be 
subject to a more elaborate law \citep[e.g., see][]{blitz_rosolowsky06}, for simplicity we adopt
a simple \citet{Ken98} law in our analytic model.}, 
with stars forming only above a certain surface gas density
threshold. Specifically, we assume that  the star formation
rate per surface area, $\dot{\Sigma}_*$, depends on the surface
density of gas, $\Sigma_{\rm g}$, as \beq \dot{\Sigma}_{*} = 2.5
\times 10^{-4} \left(\frac{\Sigma_{\rm g}}{\msol {\rm \,
pc}^{-2}}\right)^{1.4} \msol {\rm \, kpc}^{-2} {\rm yr}^{-1}\, \eeq
\citep{Ken98}.  The surface gas density profile is assumed to be
\begin{equation}
\Sigma_{\rm g} (r) = \Sigma_0 \exp\left(-\frac{r}{r_{\rm d} (z)}\right)\,.
\end{equation}
Following \cite{KGK04}, we adopt the following scaling of the disk
scale radius, $r_d$, with the total halo mass: 
\begin{equation}\label{rd}
r_{\rm d}(z) = 2^{-1/2}\lambda r_{\rm vir}(z)
\exp\left[c \left(\frac{M_4}{M}\right)^{2/3}(1+z)^{-1}\right]
\end{equation}
where $M_4 = 2.2\times 10^{10}M_\odot$ 
is the total mass of a halo with virial temperature $10^{4} K$
virializing at $z=0$, 
and $\lambda$ is the angular momentum parameter, which has a
log-normal distribution of values with scale and shape parameters 
$0.045$ and $0.56$ respectively 
\citep[e.g.,][with the peak of the distribution at 
$\lambda = 0.033$]{vitvitska_etal02}. The finite spread in possible angular 
momentum parameters  contributes to the scatter in the
observed correlations. We take $c=0.08$ by calibrating against our
simulations data. The redshift dependence in the exponential enters
through the dependence of the virial temperature on the redshift of
virialization. 

The scaling given by equation (\ref{rd}) adopts the commonly used
scaling ($r_d\propto \lambda r_{\rm vir}$) for massive halos
\citep{mo_etal98}, while for systems with the virial temperatures
$T_{\rm vir} \lesssim 10^4 {\rm K}$, it assumes that the gas cannot
cool efficiently and cannot settle into a rotationally supported disk
but settles into a more extended equilibrium configuration in the halo
\citep[see][]{KGK04}. As a result, the associated gas surface density
decreases steeply with decreasing mass and star formation is
inefficient or is completely suppressed in smaller objects. It is this inefficiency of star formation in dwarf galaxies
that is primarily responsible for the observed correlations.

Star formation is suppressed for gas surface densities $\Sigma_{\rm
  g}> \Sigma_{\rm th}\approx 5 \, \msol {\rm pc^{-2}}$
\citep{martin_kennicutt01}, which implies the threshold radius of
\beq\label{thres} r_{\rm th}(z) = r_{\rm d}(z) \ln
\frac{\Sigma_0(z)}{\Sigma_{\rm th}}\,.  \eeq This radius decreases
with decreasing galaxy mass. In other words, star formation proceeds
within smaller radii in galaxies of smaller masses.

The star formation rate $\psi $ can then be calculated as follows:
\begin{eqnarray}\label{schm}
\psi(\mg,z) &=& \int_0^{r_{\rm th}} \dot{\Sigma}_* 2\pi r dr \nonumber \\
&=& 8 \times 10^{-4} \msol {\, \rm yr^{-1}} 
\left(\frac{r_{\rm d}(z)}{{\rm kpc}}\right)^2 \left(\frac{\Sigma_0(z)}{\msol
{\rm \, pc^{-2}}}\right)^{1.4} \nonumber \\
&& \times \left[1 - \left(1+\frac{1.4r_{\rm th}(z)}{r_{\rm d}(z)}\right)\exp\left(
-\frac{1.4r_{\rm th}(z)}{r_{\rm d}(z)}
\right)\right]\,.
\nonumber \\
\end{eqnarray}
In turn, $\Sigma_0$ and its redshift dependence can be obtained through
\beq
\mg = \int_0^\infty 2\pi r dr \Sigma_{\rm g}(r) = 2\pi [r_{\rm d}(z)]^2 \Sigma_0(z)\,.
\eeq
where $r_{\rm d}$ is obtained as a function of $\mg$ from equation 
(\ref{rd}) assuming  $\mg \approx \Omega_{\rm b}M / \Omega_{\rm m}$.

Using equation~(\ref{acrate}) for the accretion rate and equation
(\ref{schm}) for the star formation rate, we can integrate equations
(\ref{gas}) and (\ref{diffstM}) numerically.  
As we show in \S \ref{res}, the
agreement of the resulting dependence of $\ms$ on $\mg$ 
with the relation obeyed by the simulated galaxies (see
Fig.~\ref{MgMst}) is striking, especially when one takes into account
that the values of the parameters $\lambda$ and $\Sigma_{\rm th}$ 
are either on or very close to their 
 ``most probable'' values.

\bibliographystyle{apj}
\bibliography{ms}

\end{document}